\documentclass[12pt,manuscript]{aastex}

\usepackage{graphics,graphicx}

\begin{document} 

%%%%%%%%%%%%%%%%%%%%%%%%%%%%%%%%%%%%%%%%%%%%%%%%%%%%%%%%%%%%%%%%%%%

\title{The M33 Globular Cluster System with PAndAS Data: The Last
  Outer Halo Cluster?}

\author{Robert Cockcroft}
\affil{Department of Physics and Astronomy, McMaster University, Hamilton, Ontario, L8S 4M1, Canada}
\email{cockcroft@physics.mcmaster.ca}

\author{William E. Harris}
\affil{Department of Physics and Astronomy, McMaster University, Hamilton, Ontario, L8S 4M1, Canada}
\email{harris@physics.mcmaster.ca}

\author{Annette M. N. Ferguson}
\affil{Institute for Astronomy, University of Edinburgh, Blackford
  Hill, Edinburgh, EH9 3HJ, UK}
\email{ferguson@roe.ac.uk}

\author{Avon Huxor}
\affil{H. H. Wills Physics Laboratory, Tyndall Avenue, Bristol BS8 1TL, UK}
\email{Avon.Huxor@bristol.ac.uk}

\author{Rodrigo Ibata}
\affil{Observatoire Astronomique, Université de Strasbourg, CNRS, 11, rue de l'Université, F-67000 Strasbourg, France}
\email{ibata@astro.u-strasbg.fr}

\author{Mike J. Irwin}
\affil{Institute of Astronomy, University of Cambridge, Madingley Road, Cambridge CB3 0HA, UK}
\email{mike@ast.cam.ac.uk}

\author{Alan W. McConnachie}
\affil{NRC Herzberg Institute of Astrophysics, 5071 West Saanich Road,
  Victoria, British Columbia, V9E 2E7, Canada}
\email{alan.mcconnachie@nrc-cnrc.gc.ca}

\author{Kristin A. Woodley}
\affil{Department of Physics and Astronomy, University of British
  Columbia, 6224 Agricultural Road, Vancouver, British Columbia, V6T 1Z1, Canada}
\email{kwoodley@phas.ubc.ca}

\author{Scott C. Chapman}
\affil{Institute of Astronomy, University of Cambridge, Madingley Road, Cambridge CB3 0HA, UK}
\email{schapman@ast.cam.ac.uk}

\author{Geraint F. Lewis}
\affil{Sydney Institute for Astronomy, School of Physics, University
  of Sydney, NSW 2006, Australia  }
\email{gfl@physics.usyd.edu.au}

\and

\author{Thomas H. Puzia}
\affil{Department of Astronomy and Astrophysics, Pontificia
  Universidad Católica de Chile, Av. Vicuna Mackenna 4860, 7820436 Macul, Santiago, Chile}
\email{tpuzia@gmail.com}

%%%%%%%%%%%%%%%%%%%%%%%%%%%%%%%%%%%%%%%%%%%%%%%%%%%%%%%%%%%%%%%%%%%

\begin{abstract}
We use CFHT/MegaCam data to search for outer halo star clusters 
in M33 as part of the Pan-Andromeda Archaeological
Survey (PAndAS).  This work extends
previous studies out to a projected radius of 50 kpc and covers
over 40 square degrees.  We find only one new
unambiguous star cluster in addition to the five previously
known in the M33 outer halo (10 kpc $\leq$ r $\leq$ 50 kpc).  Although we identify 2440 cluster candidates
of various degrees of confidence 
from our objective image search procedure, almost all of these are
likely background contaminants, mostly faint unresolved galaxies.  We
measure the luminosity, color and structural parameters of the new
cluster in addition to the five previously-known outer halo
clusters.  At a projected radius of 22 kpc, the new cluster is
  slightly smaller, fainter and redder than all but one of the other outer halo
  clusters, and has
  $g'\approx19.9$, $(g'-i')\approx0.6$, concentration
  parameter $c\approx1.0$, a core radius $r_{c}\approx3.5$ pc, and a half-light
  radius $r_{h}\approx5.5$ pc.  For M33 to have so few outer halo
clusters compared to M31 suggests either tidal stripping of M33's
outer halo clusters by M31, or a very different, much calmer accretion
history of M33.  

\end{abstract}

%%%%%%%%%%%%%%%%%%%%%%%%%%%%%%%%%%%%%%%%%%%%%%%%%%%%%%%%%%%%%%%%%%%

\section{Introduction and Background}
\label{intro}

Globular cluster systems (GCSs) are important tracers of galaxy
formation and evolution. For example, the substructure within a galactic halo
reveals its merger history, and globular clusters (GCs) can be used as
one tracer of such substructure (e.g., \citealt{1995MNRAS.275..429L}).  In this paper, we look at the
Triangulum Galaxy (M33)
which is the third most massive galaxy within our Local Group although it is
much less well-studied than the Milky Way (MW), the Magellanic Clouds
or Andromeda (M31).  

In addition to the well-known stream from the disrupting Sagittarius
dwarf galaxy, other evidence for
substructure within the Milky Way includes the Monoceros
ring, the Orphan stream, and other more subtle overdensities
(e.g., \citealt{1995MNRAS.277..781I, 2002ApJ...569..245N, 2006ApJ...642L.137B,
 2007ApJ...658..337B}).  M31's substructure is being revealed in more and more detail with large-scale structures of very low surface brightness, including
several arcs, shells and streams \citep{2002AJ....124.1452F, 2005ApJ...634..287I, 2007ApJ...671.1591I, 2006ApJ...648..389K, 2008AJ....135.1998R,
  2009Natur.461...66M}.  

Subdivisions within the MW GCS have been observed (e.g.,
\citealt{1978ApJ...225..357S, 2004MNRAS.355..504M}) with evidence that
at least some are the result of accretions of dwarf satellite galaxies
\citep{2003AJ....125..188B, 2004MNRAS.355..504M, 2004AJ....127.3394F}.  Certain clusters still appear to be associated
with their accreted satellites: most prominently, clusters associated with Sagittarius (e.g.,
\citealt{2000AJ....119.1760L, 2003ApJ...596L.191N,
  2003AJ....125..188B}).  The most distant GC known in the MW
is still AM-1, first discovered by \cite{1979ApJ...227L.103M} at
a galactocentric distance of $\approx$120 kpc.  In other galaxies, clusters with large galactocentric radii ($\approx$ 120 kpc)
reside in the $M_v$ - $r_h$ parameter space between Palomar-type clusters and
ultra-faint dwarfs, and this overlap is
now well-established (e.g., \citealt{2005MNRAS.360.1007H},
\citealt{2006A&A...447..877G} and \citealt{2007ApJ...658..337B}).  

Within M31, there are now over 60 known clusters with a projected radius greater
than 30 kpc \citep{2005MNRAS.360.1007H, 2006MNRAS.371.1983M,
  2007ApJ...655L..85M, 2008MNRAS.385.1989H,
  2010ApJ...717L..11M}.  Some of these distant clusters are rather unlike their MW
counterparts 
as they are found to be both more luminous and have larger sizes
\citep{2007ApJ...655L..85M}.  Most recently it has been shown that the outer halo clusters
appear to follow other substructure (streams of enhanced surface brightness), with the
probability of chance alignment less than 1$\%$
\citep{2010ApJ...717L..11M}.  \citeauthor{2010ApJ...717L..11M} conclude that the majority of these
clusters are accreted along with their host
satellite galaxy, as first proposed by \cite{1978ApJ...225..357S}.

Observations of the M33 clusters - both young and old - have been collated in the catalogue by
\citeauthor{2007AJ....134..447S} (\citeyear{2007AJ....134..447S}; SM hereafter).  This catalogue includes cluster identifications
and data from ground-based observations \citep{1960ApJ...131..163H,
  1978A&AS...34..249M, 1982ApJS...49..405C, 1988AJ.....95..704C,
  1998AcA....48..455M}, $HST$ imaging \citep{1999ApJS..122..431C,
    2001A&A...366..498C, 2005A&A...444..831B, 2007AJ....134.2168P,
    2007AJ....133..290S, 2008AJ....135.1482S, 2009ApJ...698L..77H, 2009ApJ...699..839S}, and further data on identified clusters from
  \cite{2001AJ....122.1796M, 2002ChJAA...2..197M, 2002AJ....123.3141M,
    2002AcA....52..453M, 2004ChJAA...4..125M, 2004A&A...413..563M}.  The SM catalogue contains 595 objects of
which 428 are classified as high-confidence clusters (based on
{\it HST} and high-resolution ground-based imaging).  The most recent work, currently not within the SM catalogue, includes work based on CFHT/MegaCam imaging by
\cite{2008AcA....58...23Z} and \cite{2010arXiv1007.1042S} and
$HST$ imaging by \cite{2009AcA....59...47Z}, that contain 3554, 599 and 91 new
star cluster candidates, respectively.  (All of these M33 studies cover only the inner one square degree.)  \cite{2009AcA....59...47Z}
claim that $\approx$20$\%$ of the 3554 cluster candidates identified
in \cite{2008AcA....58...23Z} are likely to be genuine clusters.  Unlike the GCSs of the MW and M31, M33 is host to intermediate-age
clusters \citep{1998ApJ...508L..37S, 2002ApJ...564..712C}, suggesting
that the evolution of M33 was different from that of both the MW or M31.

Studying the Local Group gives us the best chance to observe the
remnants of galaxy formation in detail, but M33 remains to be
scrutinized in as much detail as either of its larger neighboring
galaxies, or the Magellanic Clouds.  The work on the M33 GCS has so far been constrained to the classical
disk regions, with the exception of the four outer halo clusters found
by \cite{2009ApJ...698L..77H} between projected radii of 9.6 and 28.5 kpc and one cluster by \cite{2008AJ....135.1482S} at a
projected radius of
12.5 kpc.  The outer halo clusters are important, not least because
the most distant clusters may be the last that were
accreted (e.g., \citealt{2005MNRAS.360..631M}).
\cite{2010ApJ...717L..11M} have shown that M31's outer halo is rich
with clusters.  \cite{2009ApJ...698L..77H} undertook a search for M33
outer halo clusters through 12 sq. degrees of the Isaac Newton
Telescope Wide-Field Camera data reaching to
V$\sim$24.5 and i$\sim$23.5.  The PAndAS data allow this search to be extended to
larger radii, deeper depths and better image quality.  This is the
project that we undertake in this paper.  We define outer halo
clusters to be those which are projected beyond the isophotal radius
of M33 ($\sim$9 kpc, \citealt{2011Cockcroftinprep}).  Such objects are sufficiently remote that they are unlikely
to be associated with the main disk component of the galaxy: \cite{2010ApJ...723.1038M} find little evidence from direct stellar photometry
that the disk extends beyond that point.  For comparison, the
isophotal radius of NGC 253, an Sc-type galaxy of similar size,
is $r \sim 9.8$ kpc \citep{2003AJ....125..525J}.  Ultimately however, we
will require metallicity and velocity measurements to determine more
definitely whether these clusters belong dynamically to the disk or halo.

%%%%%%%%%%%%%%%%%%%%%%%%%%%%%%%%%%%%%%%%%%%%%%%%%%%%%%%%%%%%%%%%%%%

\section{Observations, Data Reduction and Calibration}
\label{obs}
We use 41 images each in $g'$ and $i'$ that are part of the
Pan-Andromeda Archaeological Survey (PAndAS; \citealt{2009Natur.461...66M}) and were taken with the Canada-France-Hawaii Telescope 
(CFHT)/MegaCam which has a one square degree field-of-view.  PAndAS includes
over 300 sq. degrees, covering a region of sky that extends to a
projected radius of 150 and 50 kpc around the Andromeda (M31) and
Triangulum (M33) Galaxies, respectively.  

Each of the 82 processed
fields, themselves a stack of four or five raw images, is labelled according to the notation shown in
Figure \ref{locations}.  Fields M72 to M76, 
the five located along a line towards M31,  were taken first \citep{2007ApJ...671.1591I}; 
we note that for these, CCD chip
4 was only working for M72, but not M73 to M76. The other fields 
M3301 to M3335 were taken in subsequent runs.  
The central image (called M33c) was a composite of CADC
archived images prepared by \cite{2007ApJ...671.1591I}. 

All M33 images were taken with sub-arcsecond seeing in both $g'$ and
$i'$.  The average seeing on $g'$ frames was 0.75'' (standard
deviation of 0.11''), and 0.66'' on $i'$ (standard deviation of
0.13'').  Images have a resolution of 0.187''/pixel, and
limiting magnitudes of $g' \approx$ 25.5, $i' \approx$ 24.5 (AB mags on
the SDSS scale) at an
S/N = 10.  These data were previously presented in
\cite{2010ApJ...723.1038M} who studied the stellar structure of the
outer regions of M33 and found a large substructure, shown in Figure \ref{locations}.  The
data were pre-processed with Elixir\footnote{http://www.cfht.hawaii.edu/Instruments/Imaging/MegaPrime/dataprocessing.html} by the CFHT team, and then reduced
at the Cambridge Astronomical Survey
Unit through a pipeline adapted for
MegaCam images \citep{2001NewAR..45..105I}.  

%%%%%%%%%%%%%%%%%%%%%%%%%%%%%%%%%%%%%%%%%%%%%%%%%%%%%%%%%%%%%%%%%%%

\section{Cluster Search Methods}
\label{search}

We used two distinct methods to search for clusters within the M33
images: an automatic search and a separate follow-up visual inspection.  In both cases, we started with searching the images that form
an annulus in the middle of the frames around M33 (i.e., frames M3306 to M3316, including
M74), followed by the images outside this annulus
and finally the innermost frames.  We chose this order so that we started in regions where the
crowding was low but evidence for clusters existed 
(e.g., \citealt{2009ApJ...698L..77H}), leaving the innermost crowded
fields until last.  We identified and marked all the confirmed clusters 
in the \cite{2007AJ....134..447S} catalogue on the PAndAS images, 
to gain experience of their appearance.

\subsection{Automatic Search}
\label{auto}

We used Source Extractor (SE; \citealt{1996A&AS..117..393B}) to
identify all objects in both $g'$ and $i'$ frames.  Some values within
the configuration files were changed to optimize finding clusters
whilst cutting out as much contamination as possible (see Table
\ref{separams}).  We then converted from a chip-oriented pixel
coordinate system to a world coordinate system for each object, using
the IRAF/wcsctran routine, before matching the objects across the $g'$ and $i'$
frames using the IRAF/xyxymatch routine.  As the latter routine only
outputs object coordinates, we then re-assigned all SE parameters to
the matched objects, so that we could apply selection
criteria using a combination of magnitude, color, half-light radius, and 
ellipticity as measured by SE to pick out the cluster candidates.  
After numerous initial tests and iterations, we have 
adopted the following set of criteria:
\begin{equation}
\label{eqn1}
10.5 \leq g' \leq 14.5,
\end{equation}

\begin{equation}
\label{eqn2}
-1.1 \leq g'-i' \leq -0.175*g'+3.4375,
\end{equation}

\begin{equation}
\label{eqn3}
e \leq 0.375, 
\end{equation}

\begin{equation}
\label{eqn4}
3.5 \leq r_{flux} \leq 16.0, 
\end{equation}
and
\begin{equation}
\label{eqn5}
r_{flux} \leq -2.125*g'+41.5, 
\end{equation}

\noindent where $e$ is the ellipticity(1 - minor/major), $g'$ and $i'$ are the automatic
magnitude values returned from SE\footnote{The following are
  approximate conversions between true color-corrected magnitudes and
  SE automatic magitudes: $g'_{true}$ = $g'_{SE}$ + 6.2 and
  $i'_{true}$ = $i'_{SE}$ + 6.4.}, and r$_{flux}$ is the radius (in pixels)
estimated to enclose half the flux.  The cluster candidates
we select satisfy \emph{all} five of the criteria.

\begin{figure}[!htb]
  \begin{center}
    \includegraphics[width=80mm]{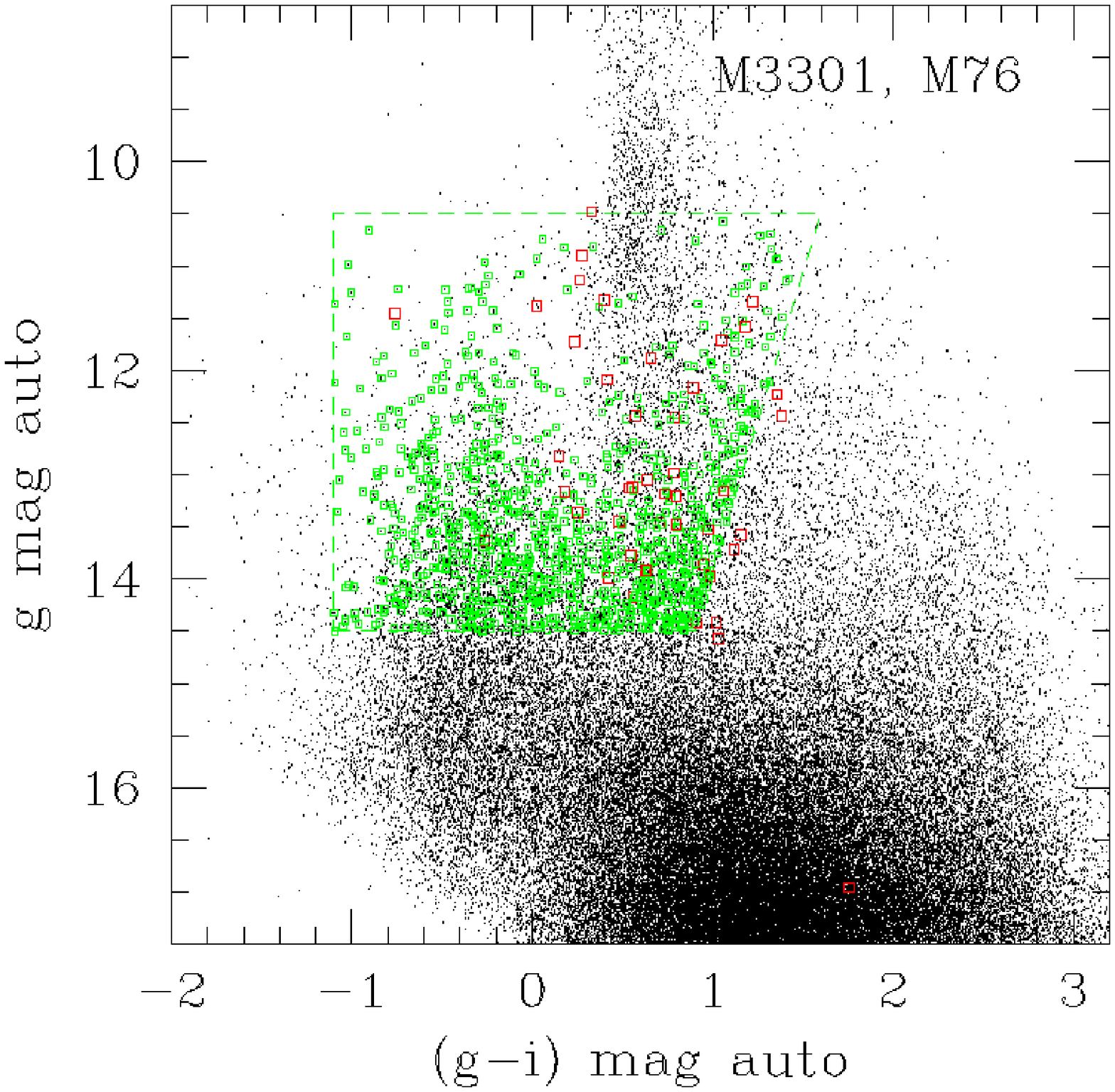}
    \includegraphics[width=80mm]{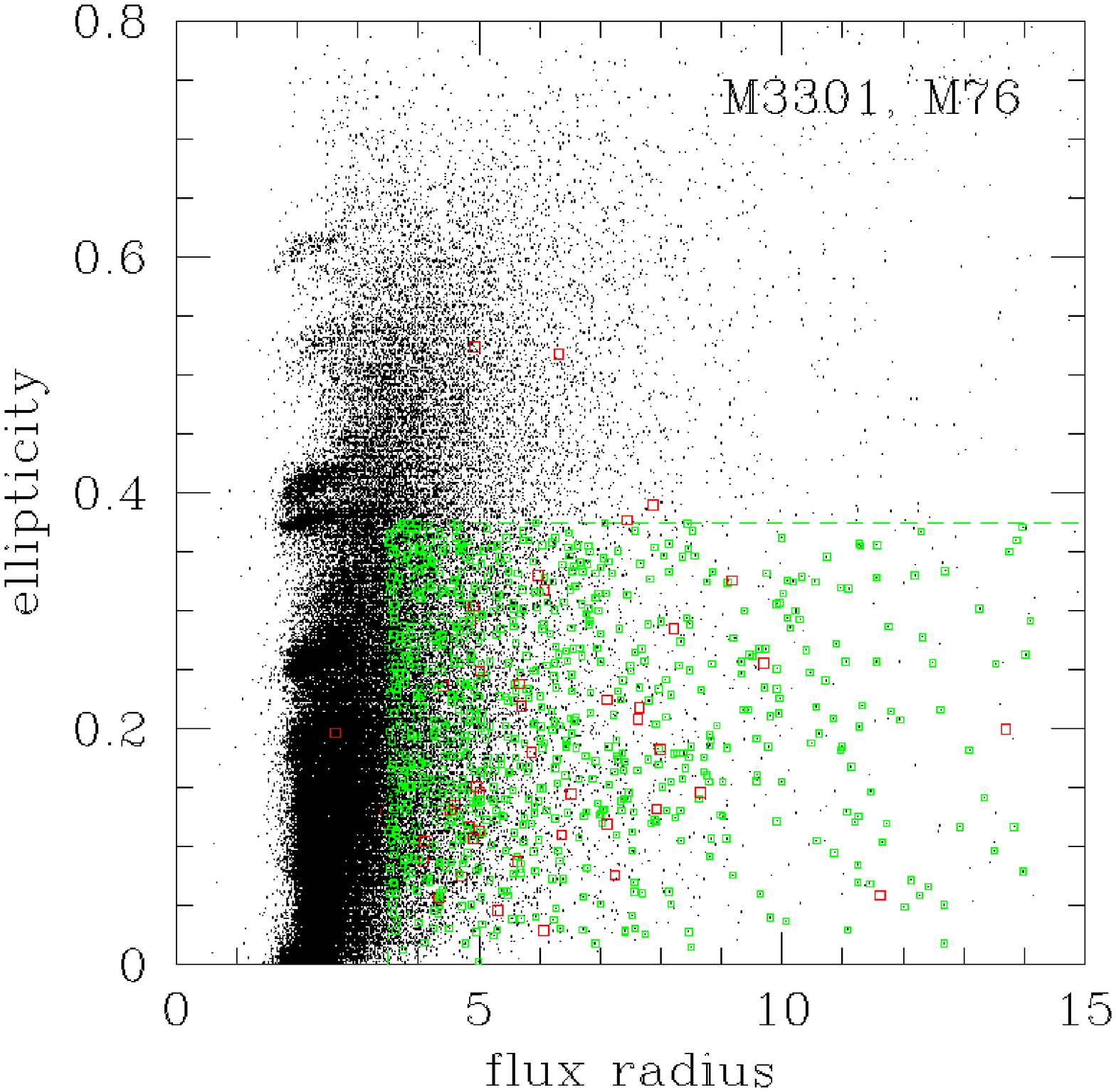}
    \includegraphics[width=80mm]{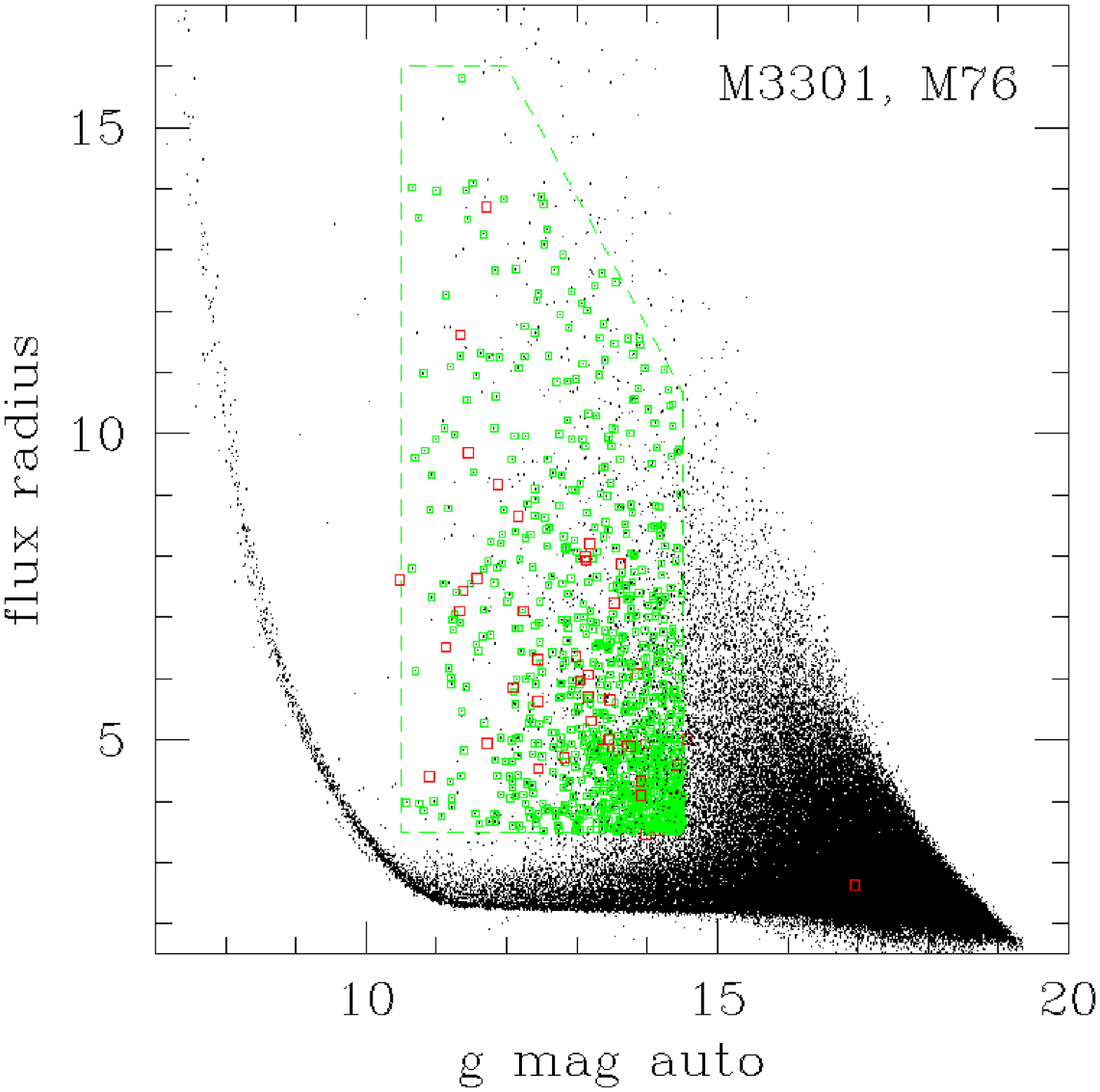}
  \end{center}
  \caption{Two halo fields and objects on which Section \ref{auto}'s selection
    criteria were based.  The black dots are the  objects of all kinds 
detected 
  by SE.  The green dashed lines show the boundaries of the selection criteria
  (equations \ref{eqn1} to \ref{eqn5}), and the green squares enclose those points which were picked
  out by all five of the selection criteria.  The red squares show the
  high-confidence clusters in the \cite{2007AJ....134..447S}
  catalogue.  The following are
  approximate conversions between true color-corrected magnitudes and
  SE automatic magitudes: $g'_{true}$ = $g'_{SE}$ + 6.2 and
  $i'_{true}$ = $i'_{SE}$ + 6.4.}
  \label{selection}
\end{figure}

\begin{figure}
  \begin{center}
    \includegraphics[width=80mm]{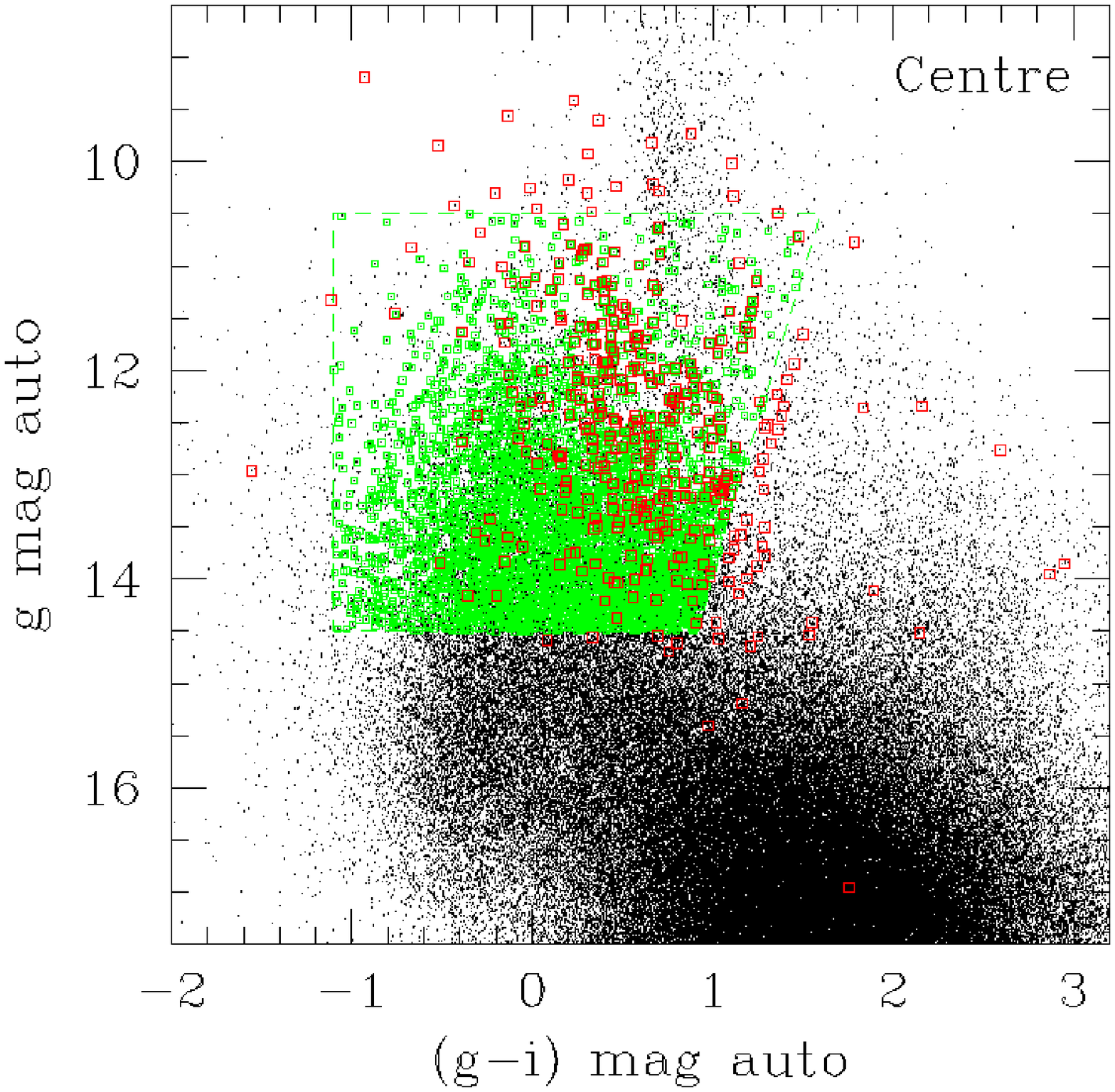}
    \includegraphics[width=80mm]{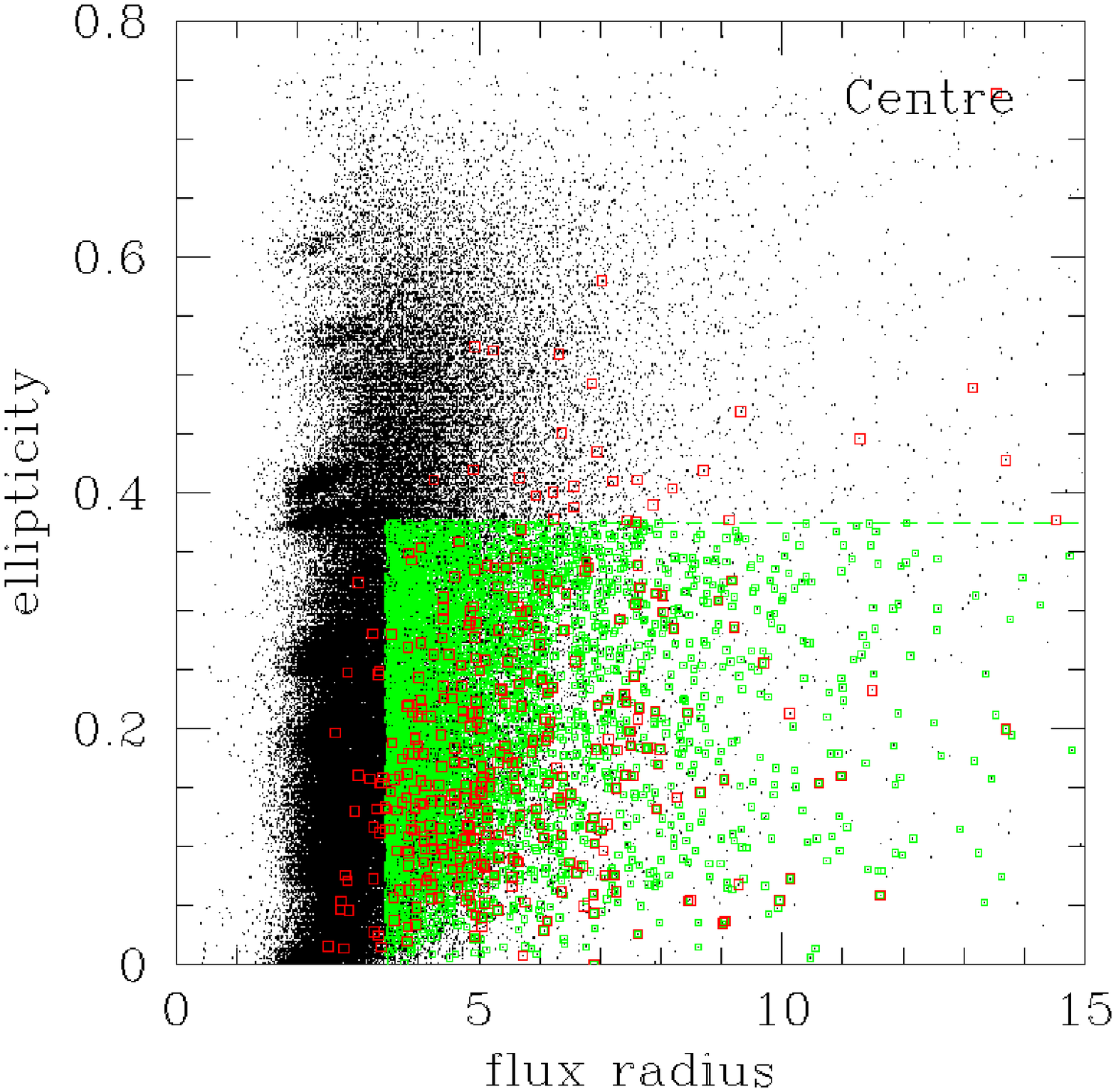}
    \includegraphics[width=80mm]{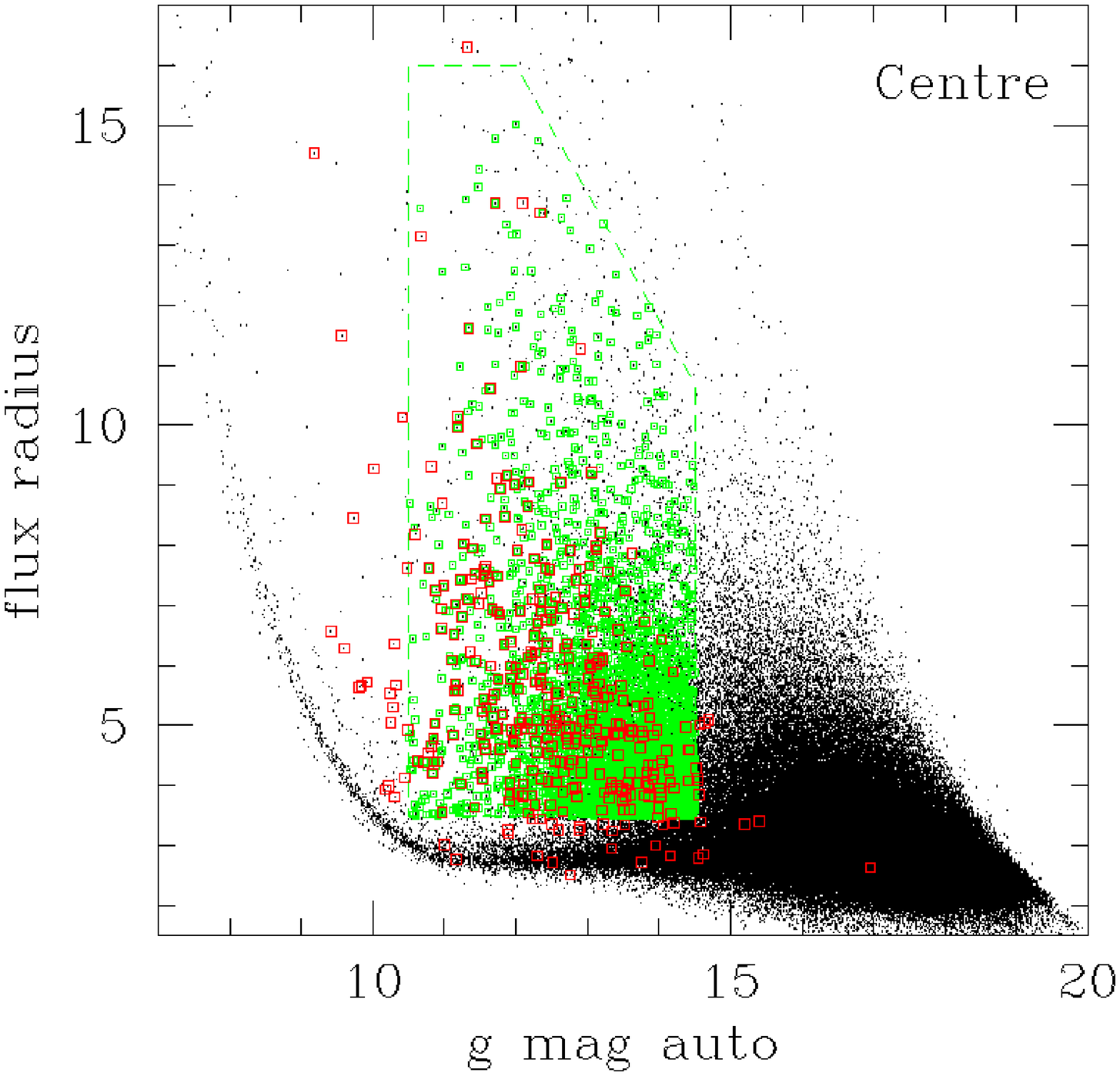}
  \end{center}
  \caption{Objects within the central field.  As with Figure
    \ref{selection}, the black dots show the SE detections.  The green dashed lines show the selection criteria, and the green squares enclose those points which were picked
  out by all five of the selection criteria.  The red squares show the
  high-confidence clusters in the \cite{2007AJ....134..447S}
  catalogue.  The following are
  approximate conversions between true color-corrected magnitudes and
  SE automatic magitudes: $g'_{true}$ = $g'_{SE}$ + 6.2 and
  $i'_{true}$ = $i'_{SE}$ + 6.4.}
  \label{all}
\end{figure}

These selection criteria are shown in Figure \ref{selection}.  The
boundary lines 
were chosen so that they included almost all of the
\cite{2007AJ....134..447S} catalogue confirmed clusters at the
edge of the disk (in the MegaCam fields M3301 and M76), while cutting
out most of the contaminating objects such as stars and background
galaxies.  The central field, however, provides a special
challenge because of the complex structure of the background light
and differential reddening.  As can be seen in Figure \ref{all},
our parameter boundaries do not include every one of the
\citeauthor{2007AJ....134..447S} high confidence clusters (see Section
\ref{intro}) in the central field.

However, our aims here were specifically to isolate candidate clusters
in the halo regions.  Other types of objects (especially background
galaxies) populate all areas of the three parametric diagrams in
Figures \ref{selection} and \ref{all}, and after many iterations we adopted the
boundary lines shown as a compromise between excluding contaminants
and including real clusters.  Nevertheless, the unavoidable fact is that
our survey area is so large (more than 40 square degrees around
M33) that even our most careful objective search criteria leave
a very large number of field contaminants, which dominate the
numbers of objects found in the range of magnitudes, colors, and
sizes that we are looking for.

\begin{figure}
  \begin{center}
    \includegraphics[width=25mm]{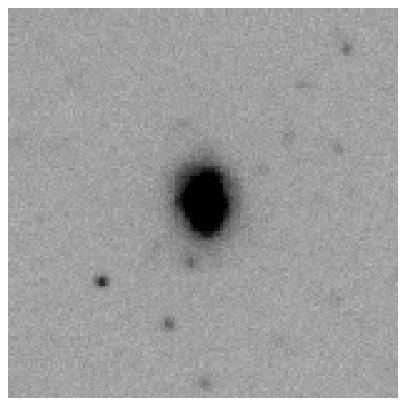}
    \includegraphics[width=25mm]{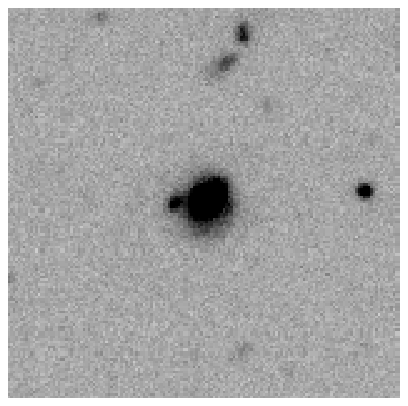}
    \includegraphics[width=25mm]{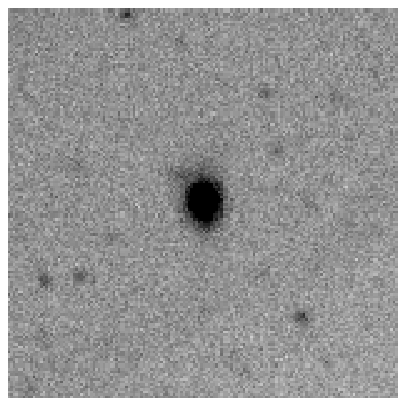}
    \includegraphics[width=25mm]{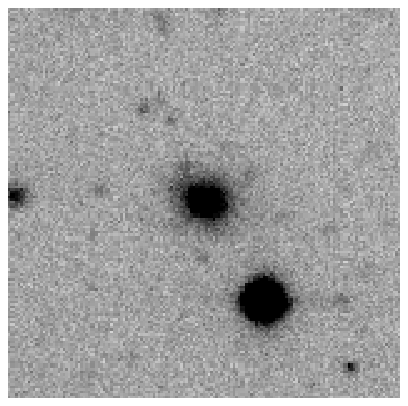}
    \includegraphics[width=25mm]{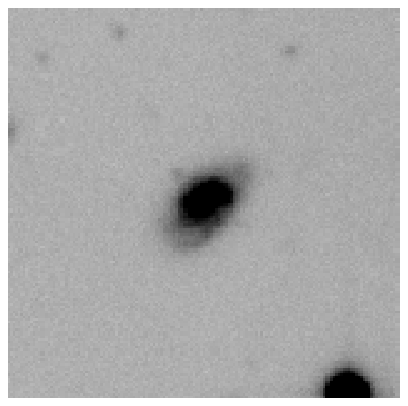}
    \includegraphics[width=25mm]{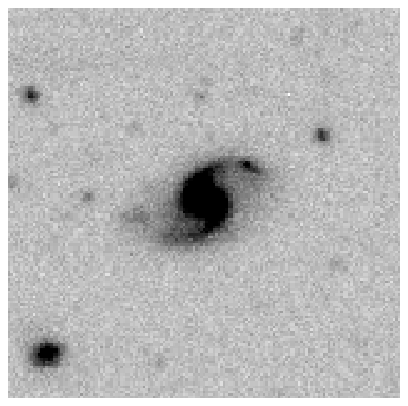}
    \includegraphics[width=25mm]{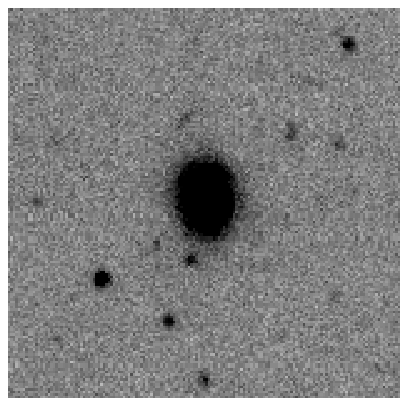}
    \includegraphics[width=25mm]{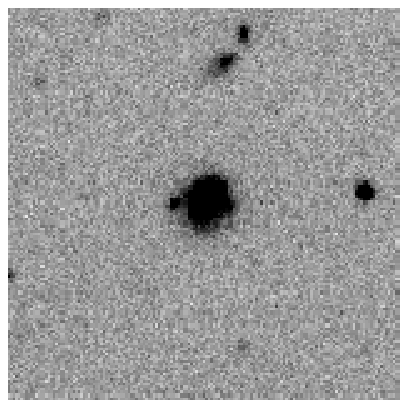}
    \includegraphics[width=25mm]{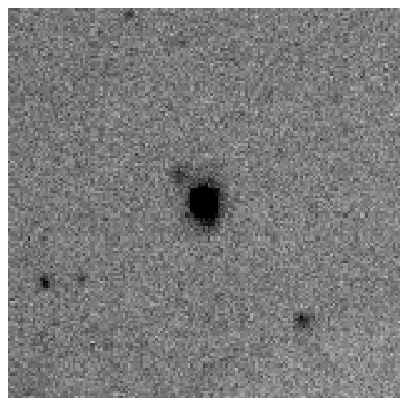}
    \includegraphics[width=25mm]{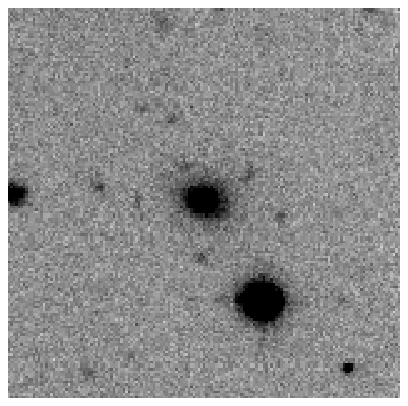}
    \includegraphics[width=25mm]{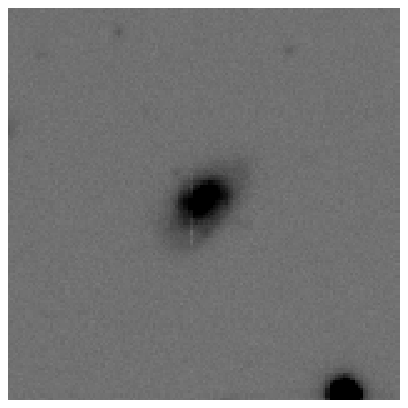}
    \includegraphics[width=25mm]{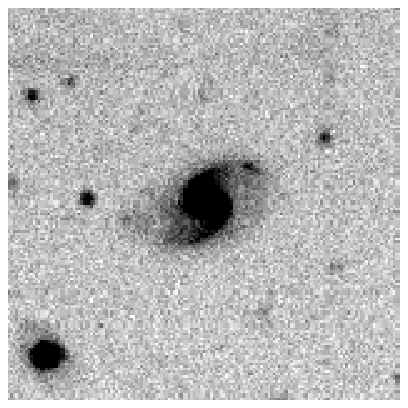}
  \end{center}
  \caption{Six examples of classified objects in $g'$ (top row) and $i'
 $ (bottom).  From left to right: two examples each of 1 (high-confidence
    clusters), 2 (possible cluster candidates), and 3 (galaxies).
    More detail is apparent when changing the scale and contrast in a
    DS9 window.  Each box is 20'' square, corresponding to about 84 pc square (at
    870 kpc).}
  \label{examples_of_classifications}
\end{figure}

We produced a small thumbnail display region 
to identify each object selected by the above
criteria.  These regions were then displayed in the $g'$ frame.  
Each object within these regions was then inspected visually and
classified following the description in Section \ref{visual} with 1 (high confidence cluster), 2 (possible cluster), 3
(background galaxy), 4 (unknown object) or 5 (stellar
object).  Examples of objects in the categories 1, 2 and 3 are shown
in Figure \ref{examples_of_classifications}.

\subsection{Visual Inspection}
\label{visual}

The next stage in our classification
procedure was, following Section \ref{auto}, to inspect all objects
that had $not$ already been selected by the automated criteria, i.e.,
we looked at all SE detections that did not fall within the selection
boxes shown in Figure \ref{selection}.  This was similar to the method
employed by \cite{2008MNRAS.385.1989H, 2009ApJ...698L..77H}.  Only
$i'$ frames were inspected this way; red-giant branch stars are 
brighter in $i'$ than in $g'$ and so clusters, and RGB stars in their halo,
would appear more obvious by their resolution into stars (no
background galaxies would be resolved into individual stars).  By also
conducting a visual search in addition to the search via the selection
criteria, we ensured that any obvious cluster or candidate cluster
would not be overlooked, and also ensured that all $g'$ and
$i'$ frames will have been inspected. 

The easiest objects to classify were the obvious clusters and background galaxies.  Clearly-resolved clusters appeared as having a circular or
slightly elliptical core, with uneven contours and resolved stars
around the central core.  Less obvious were group 1 objects which had
slightly uneven contours.  The least obvious candidates, group 2
objects, were the compact objects that could be clusters or
galaxies, and as a result the numbers of ``possible clusters'' were
the greatest especially 
in the central regions.  As the contrast and scaling were changed, some
objects smoothly grew, some were a faint smear with no sharp edges,
and others could be seen to display spiral shape.
If the object had smooth contours and if it was in a
group of other objects that were clearly galaxies, the object was
likely to be a galaxy and not a star cluster. Group 4 objects did not
look like a cluster, a cluster-candidate, a galaxy or a star.  

As noted above, the influence of background contamination by galaxies on this selection
process should not be underestimated.  In essence, this is a
needles-in-a-haystack process where we are attempting to find a small number
of clusters in a huge population of contaminants, and even though
our selection and culling is rigorous, there remain a large number of
objects whose nature is ambiguous from the current data.  Higher resolution
imaging, imaging in the near infrared where the cluster red giants
would be better resolved (and which also can have better seeing), or ultimately spectroscopy, will be required
for more definitive elimination of the last contaminants.

%%%%%%%%%%%%%%%%%%%%%%%%%%%%%%%%%%%%%%%%%%%%%%%%%%%%%%%%%%%%%%%%%%%

\section{Results}
\label{results}

\begin{figure}
  \begin{center}
    \includegraphics[width=50mm]{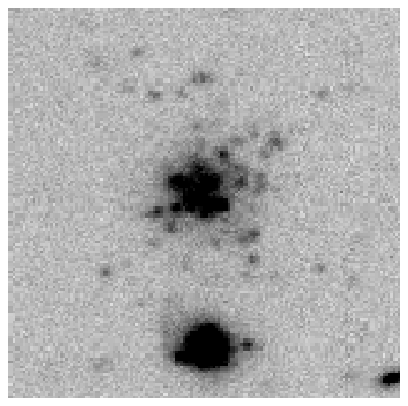}
    \includegraphics[width=50mm]{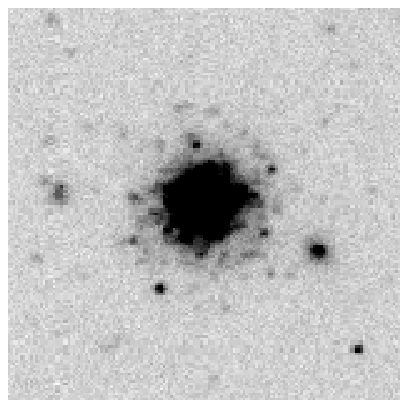}
    \includegraphics[width=50mm]{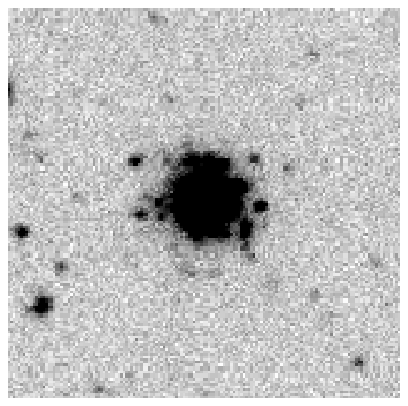}
    \includegraphics[width=50mm]{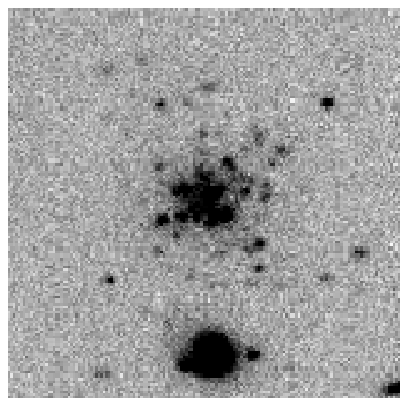}
    \includegraphics[width=50mm]{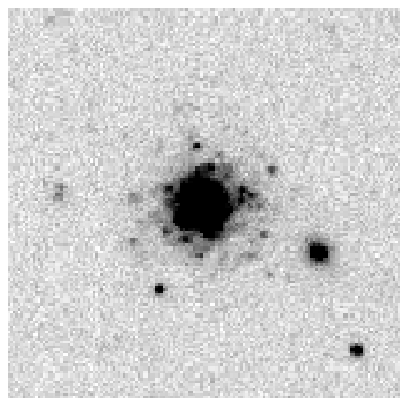}
    \includegraphics[width=50mm]{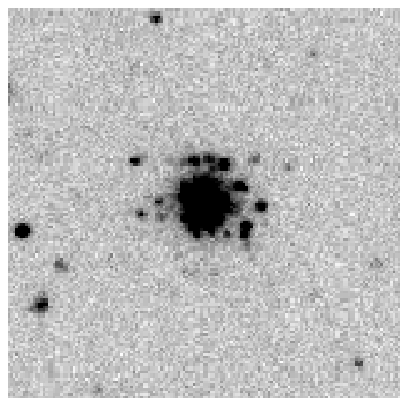}
    \includegraphics[width=50mm]{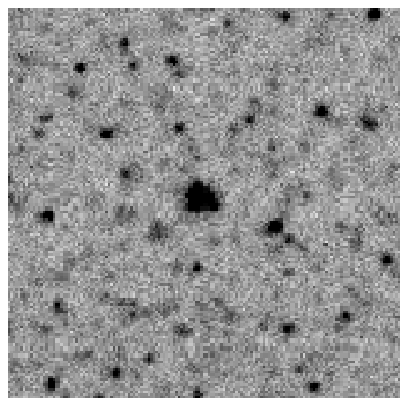}
    \includegraphics[width=50mm]{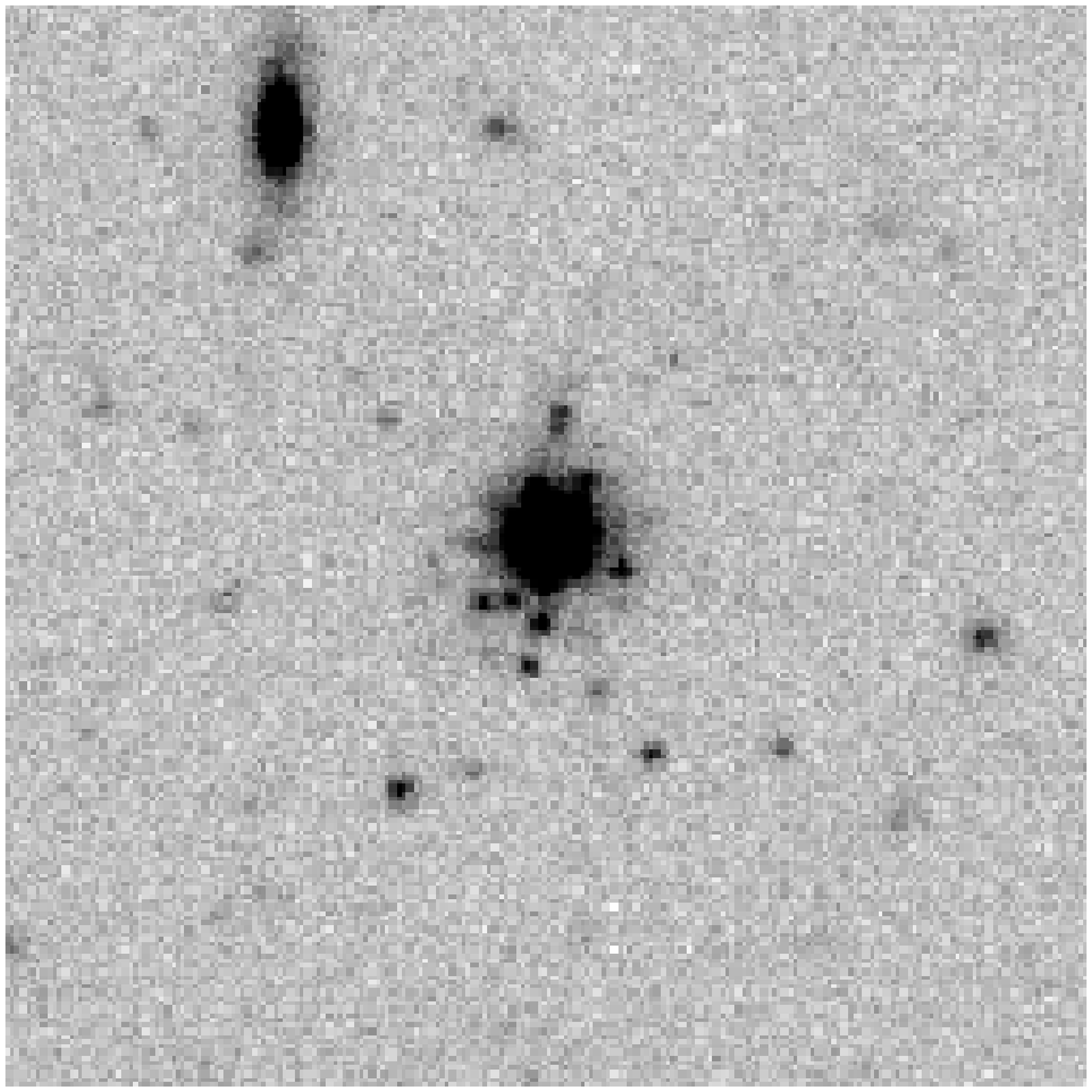}
    \includegraphics[width=50mm]{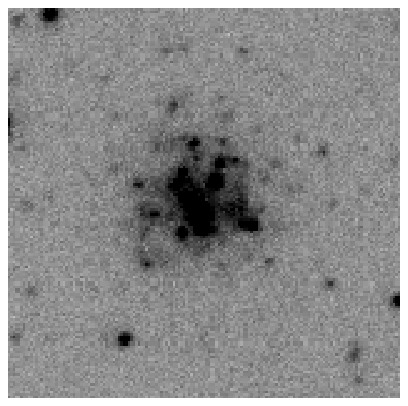}
    \includegraphics[width=50mm]{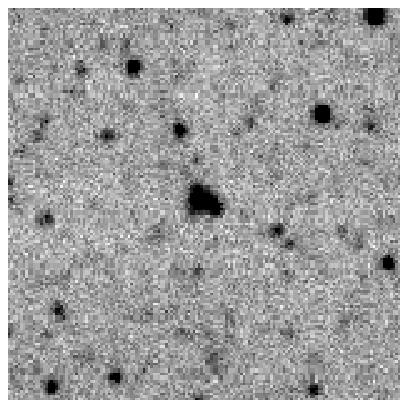}
    \includegraphics[width=50mm]{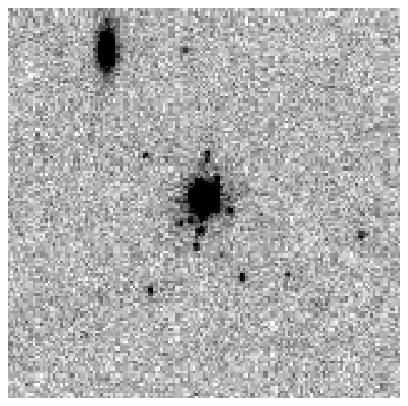}
    \includegraphics[width=50mm]{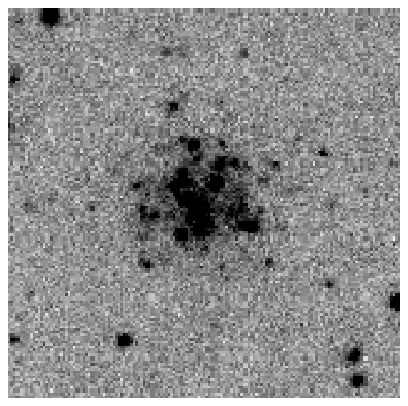}
  \end{center}
  \caption{M33 A, B, C, D, E and S (left to right) in $g'$ (top) and $i'$ (bottom).  Each box is 20'' square, corresponding to about 84 pc square (at
    870 kpc).}
  \label{m33ag}
\end{figure}

There was only one definite new
outer halo cluster discovered in our study at a projected
radius of 87'' (or 22 kpc, assuming a distance to M33 of 870 kpc).  It was found using the
automated search.  The
new cluster is named M33E following the naming convention begun in
\cite{2009ApJ...698L..77H}.  Four of the five previously-known outer halo clusters
\citep{2008AJ....135.1482S, 2009ApJ...698L..77H} were easily
recovered.  Cluster D was identified but was too compact to have been recovered without prior knowledge.  Clusters A to E and S are shown in Figure \ref{m33ag},
where S is the cluster found by \cite{2008AJ....135.1482S}.  

There were 2440
candidates spread throughout the M33 halo; that is,
in the region outside of the central MegaCam image.  87 (5)
highest-confidence cluster candidates and 2294 (54) possible 
clusters were found by the automated (visual) search method.

\begin{figure}
  \begin{center}
    \includegraphics[width=170mm]{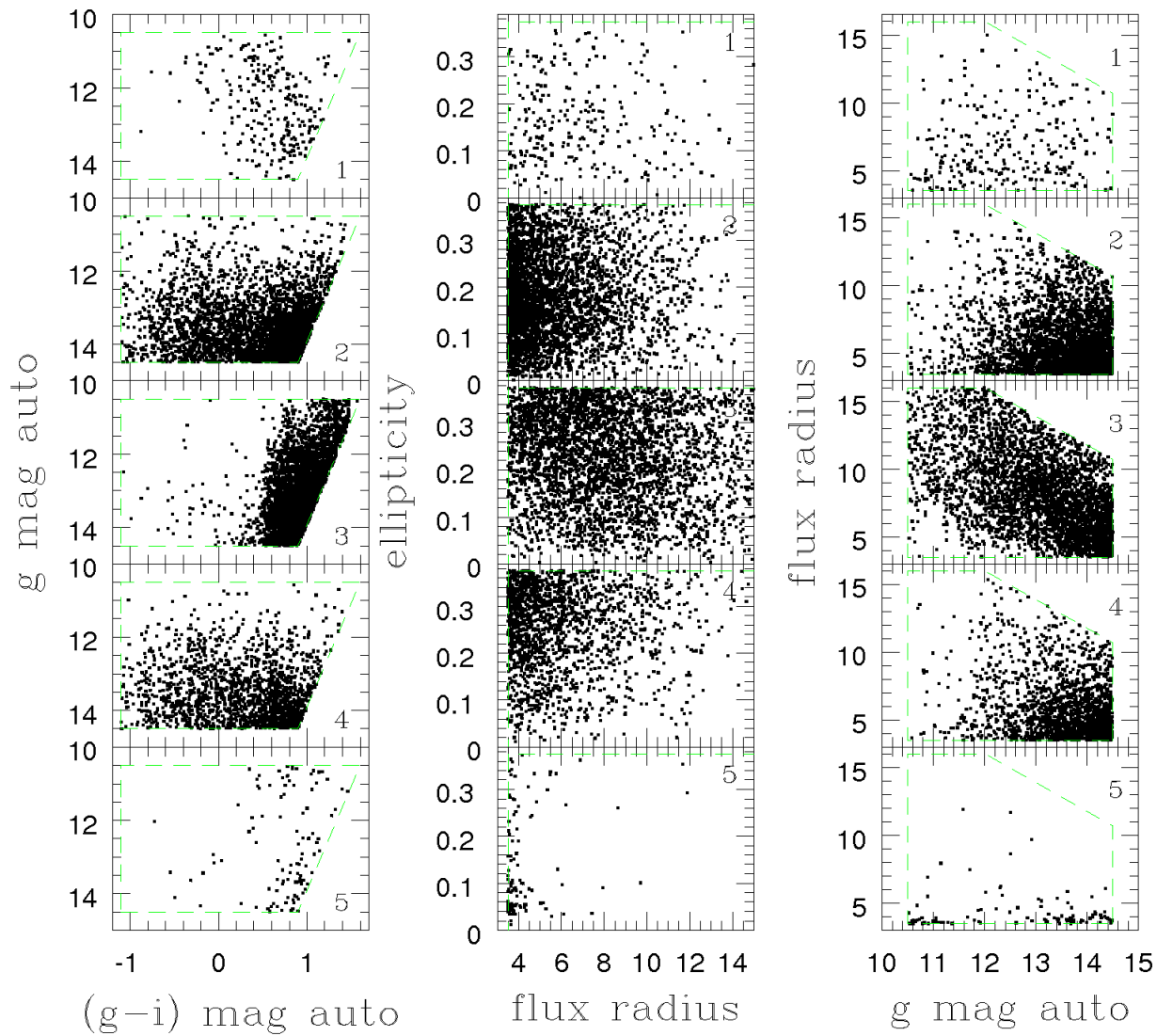}
 \end{center}
  \caption{The classified objects after visual inspection shown in
    relation to the automatic selection criteria of Section
    \ref{auto}.  Class 1 represents high confidence candidates, class 2
    are possible clusters, class 3 are background galaxies, class 4
    are unknown objects, and class 5 are stellar objects.  To
    calibrate the ``mag auto'' values, we added the zeropoint term and
    corrected for airmass and color.}
  \label{where}
\end{figure}

\begin{figure}
  \begin{center}
    \includegraphics[width=55mm]{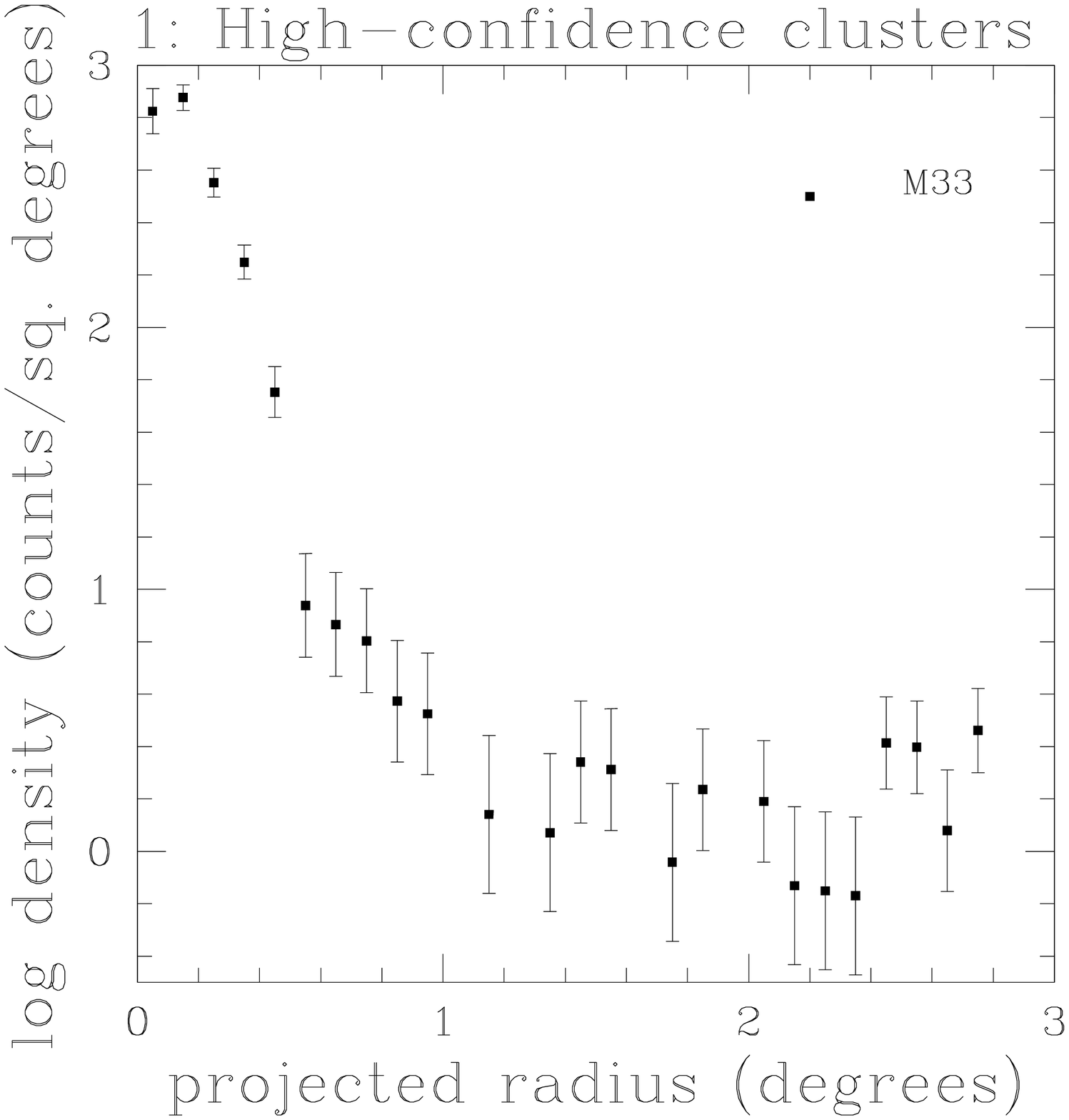}
    \includegraphics[width=55mm]{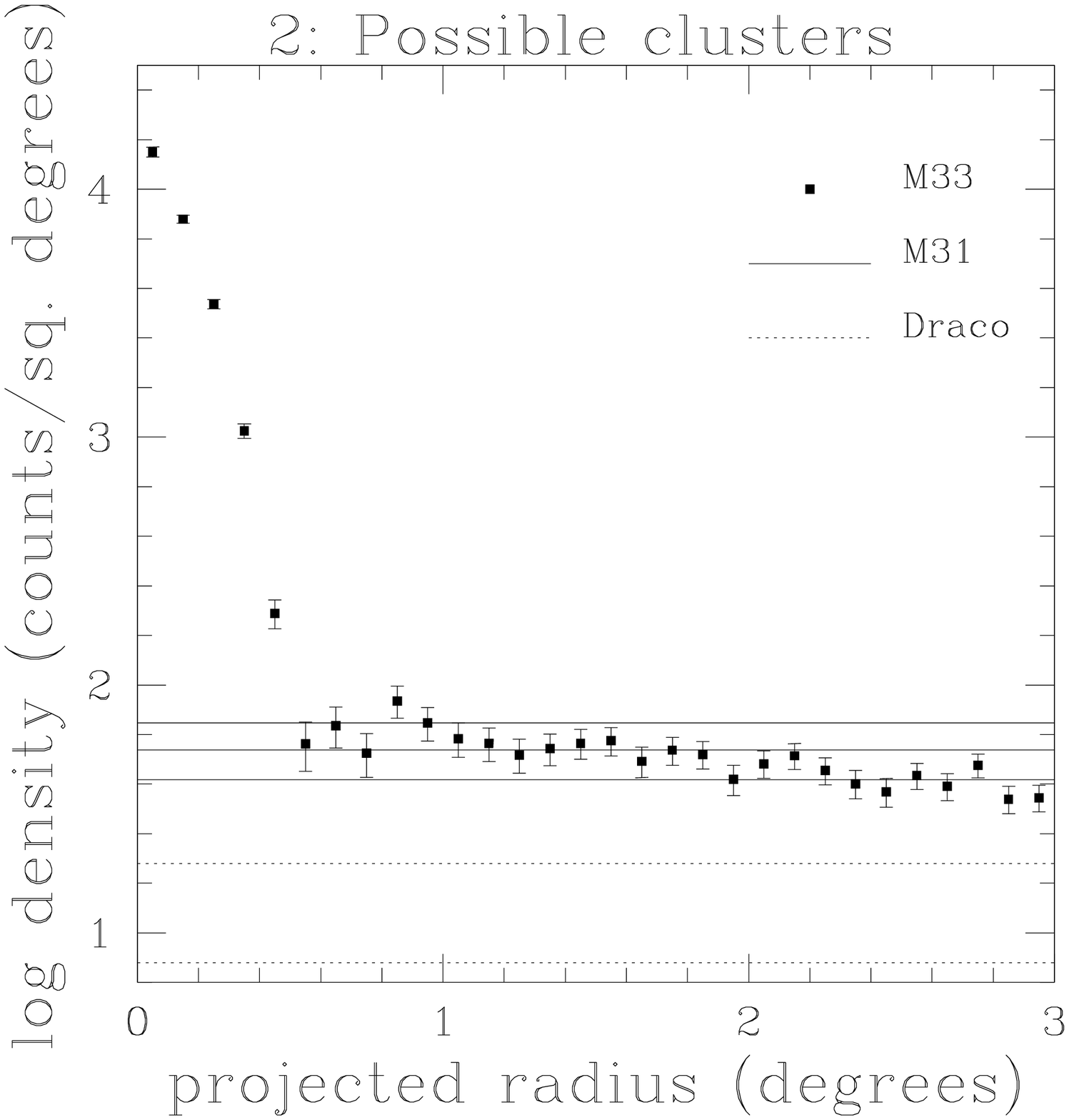}
    \includegraphics[width=55mm]{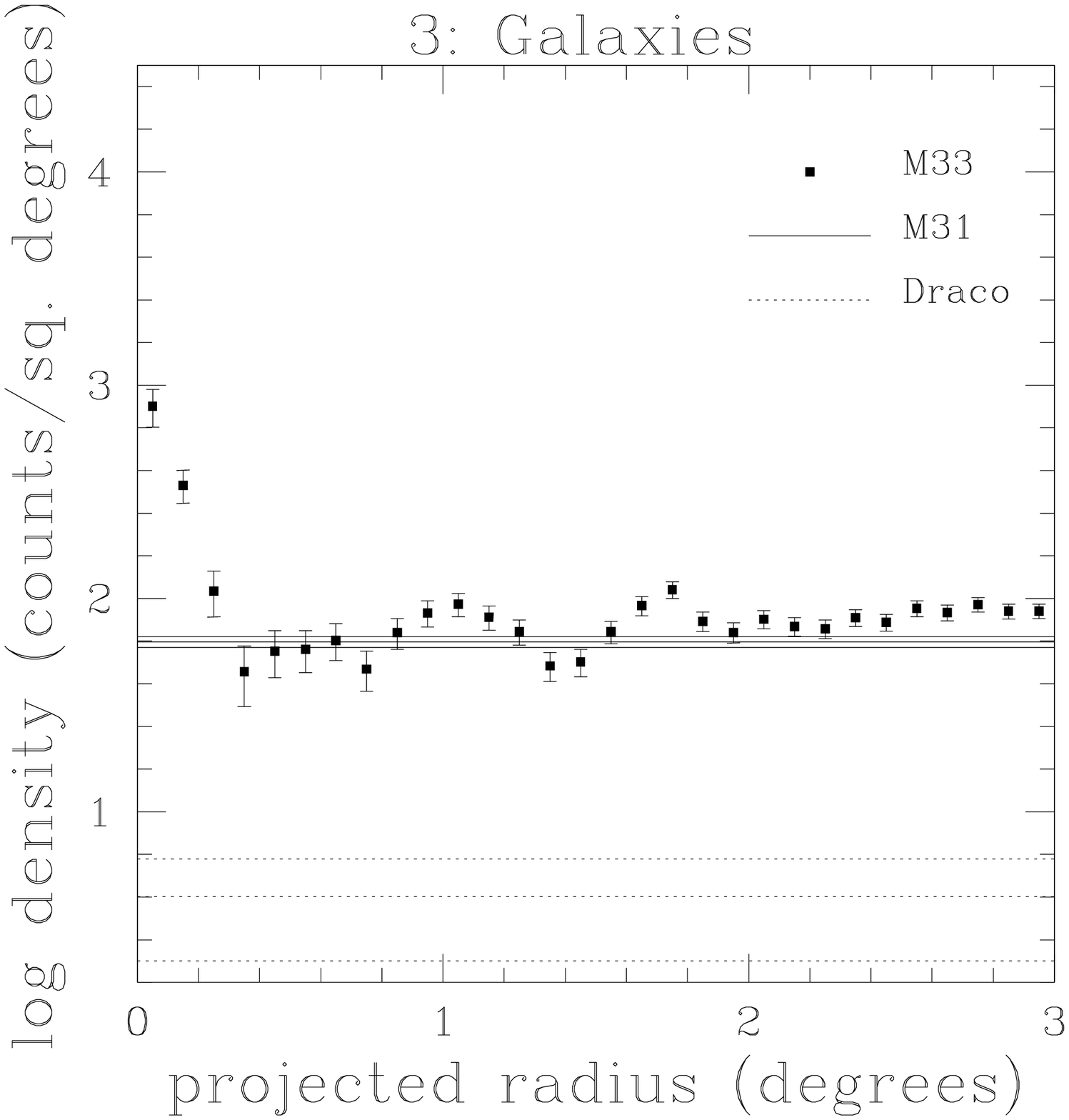}
    \includegraphics[width=55mm]{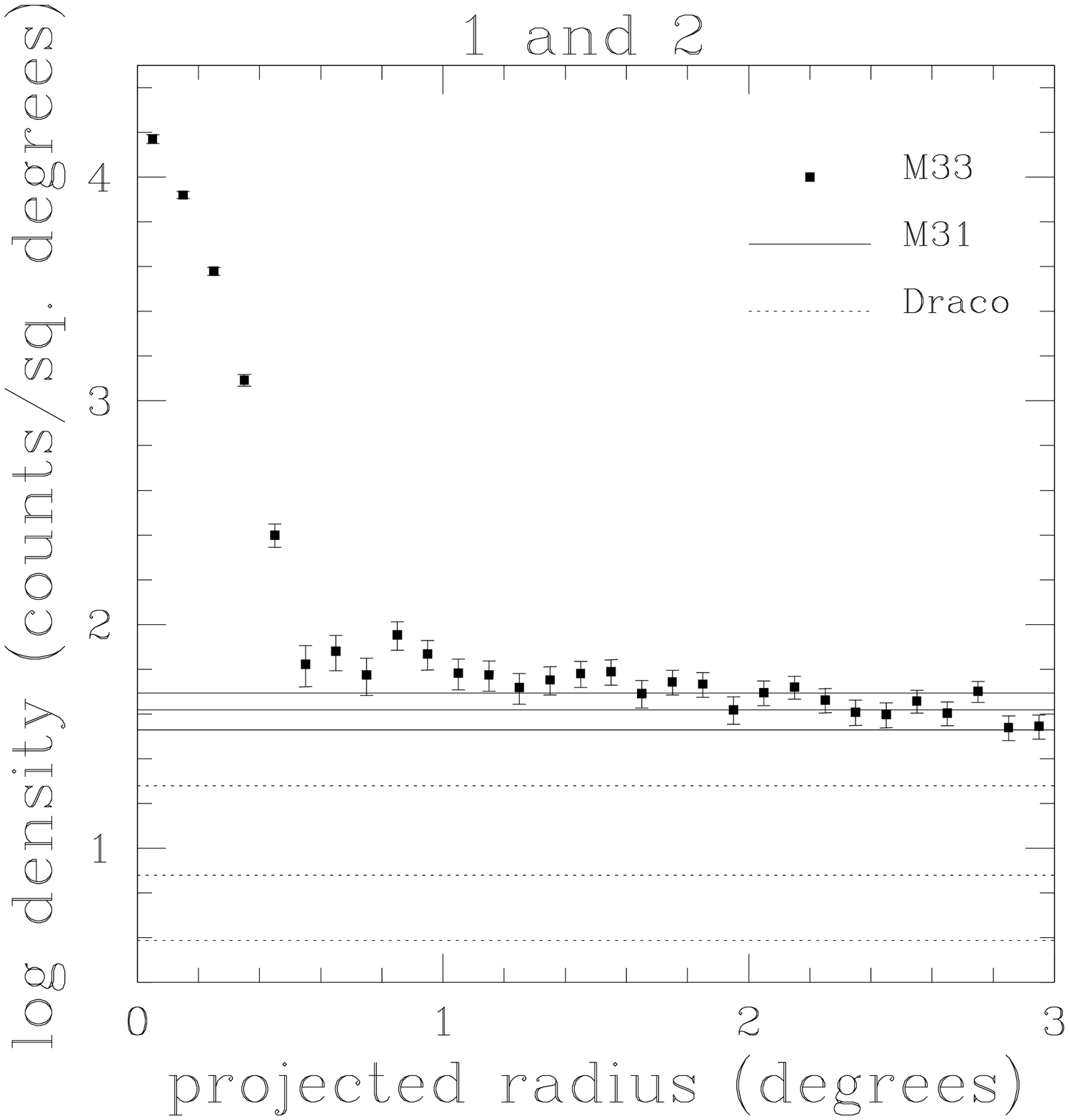}
    \includegraphics[width=55mm]{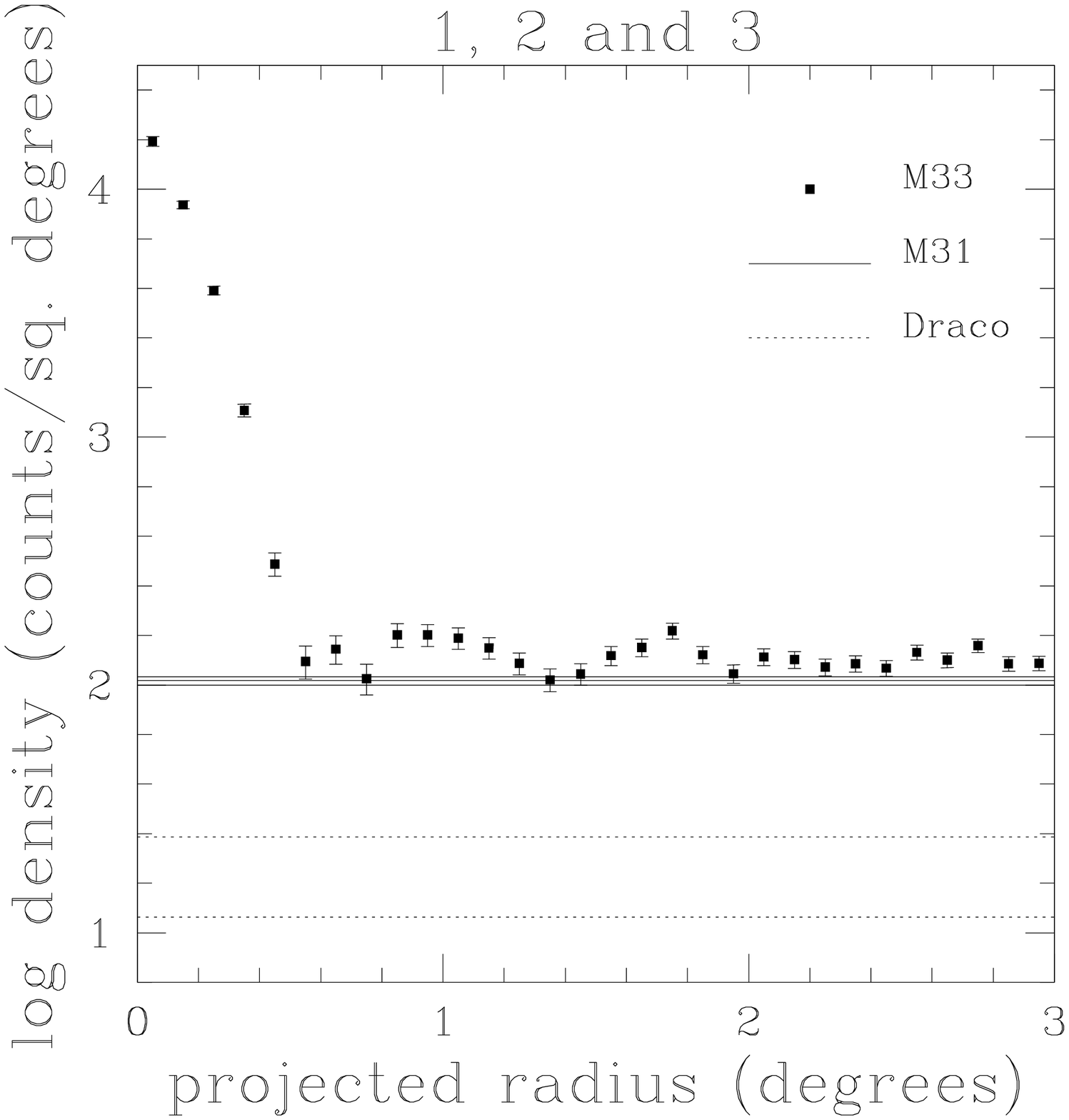}
  \end{center}
  \caption{Radial densities of objects in circular annuli from categories 1
    (high-confidence clusters), 2 (possible clusters) and 3
    (background galaxies)
    along the top, and the sum of 1 and 2, and also 1, 2 and 3 on
    the bottom.  Error bars on the points are simple $n^{1/2}$ uncertainties.  
Also shown are the mean and 1-sigma errors of the
    same categories of objects in control fields from the M31 outer
    halo (solid line) and Draco (dashed line).  The Draco number
    densities are so low, as mentioned in Section \ref{results}, that in two
    cases the 1-sigma errors are larger than the mean and so cannot be
    plotted on a log scale.  In these cases, only the mean and upper
    1-sigma error are shown.} 
  \label{rad_density}
\end{figure}

The numbers of all classified objects from both the automated and
visual inspection searches are shown in Table \ref{objects}.  
Results of the above searches were plotted within the original selection
criteria, and are shown in Figure \ref{where}.  We wanted to exclude
the maximum amount of parameter space so that we could increase the
efficiency of the automated search, and it is not obvious from this
figure that more space could have been excluded.  Radial density plots for the categories 1 (high confidence cluster), 2 (possible cluster), 3
(background galaxy) are also shown in
Figure \ref{rad_density}.  We compared these number densities at large
radii to 
control fields, also taken with MegaCam and with very
similar image quality, from the M31
outer halo and the field near the Draco dwarf spheroidal.  
The M31 fields are two square-degree fields selected directly from the
PAndAS data, at a similar Galactic latitude to M33 of -31.33 degrees, at the edge of the PAndAS footprint around M31 (i.e., at a
projected radius of $\sim$150 kpc) and did not contain any clusters -
either previously-known clusters, or clusters detected in the PAndAS images.  The Draco fields are
seven square-degree fields at a Galactic latitude of 34.72 degrees
\citep{2007MNRAS.375..831S}.  Our searches were again applied
to the control fields following exactly the same selection criteria,
and we obtained an average density of each category of objects in the
control fields.  The radial distribution plots indicate that few if any of
the category 3 objects are genuine clusters since they show little
detectable central concentration to the galaxy outside the crowded
disk region.  For all three categories plotted, the number density settles down
to a virtually constant level similar to that of the M31 control fields for
$r \gtrsim 1$ degree, consistent with the conclusion that there are few
clusters left to be found in the M33 halo down to the PAndAS limiting
magnitudes.  (The number density of all objects in the Draco fields
is much lower than that in either the M33 or M31 fields, highlighting that it
was appropriate to compare M33 with the M31 control fields.)  If we count the number of candidates for the combined objects of
classes 1 and
2 in Figure \ref{rad_density} for $r\geq$ 10 kpc, and then subtract off the
M31 background, we are left with approximately 210$\pm$130 candidates
(the error is estimated using the error on the M31 background). This
number is simply an estimate of the outer halo clusters that possibly
remain to be discovered, using the data we have in hand.  210 clusters
would be a generous upper limit, given the field contamination issues
that we discuss.

We next measured the $g'$ and $i'$ magnitudes, and the colors of the six outer halo
clusters.  The results are shown in Table \ref{clusterlums}.  For
clusters A, B, C and S we use an aperture radius of 40 pixels (7.5'')
and a sky annulus between 60 and 80 pixels (11.2''-15.0''); for the smaller
clusters we used 20 pixels (3.7'') with a sky annulus between 20
and 40 pixels (3.7''-7.5'') for D, and 30 pixels
(5.6'') with a sky annulus between 50 and 70 pixels (9.4''-13.1'') for E.  We
assume an extinction correction of 0.16 in $g'$, 0.09 in $i'$, 0.14 in $V$,
and 0.08 in $I$ \citep{1998ApJ...500..525S}, and a distance of 870 kpc to M33, consistent with SM07
and \cite{2009ApJ...698L..77H}, and corresponding to a distance
modulus of (m-M)$_0$ = 24.69.  We note that there is some
  disagreement in the literature regarding the distance to M33; see
  references in \cite{2010ApJ...723.1038M}.  For the magnitude and
  color conversions from $(g',i')$ to $(V,I)$ we used 

\begin{equation}
\label{eqn6}
V = g - (0.587 \pm 0.022)(g-r) - (0.011 \pm 0.013)
\end{equation}

\noindent and

\begin{equation}
\label{eqn7}
I = i - (0.337 \pm 0.191)(r-i) - (0.370 \pm 0.041)
\end{equation}

\noindent from \cite{2008AJ....135..264C}, and 

\begin{equation}
\label{eqn8}
(r-i)_0 = 0.37(g-r)_0 + 0.006
\end{equation}

\noindent from \cite{2008MNRAS.384.1178B}.

Comparing the magnitudes and colors that we measure in this paper with
those in 
\cite{2009ApJ...698L..77H} for clusters A, B, C and D, we find
differences of $V_0$$\leq$ 0.3mag and $(V-I)_0\leq$ 0.1mag.  For
cluster S, \cite{2008AJ....135.1482S} measures a $V$ magnitude of 18.5 (at $\sim2r_h$, which roughly
corresponds to our annulus size).  As we measure $V_0$$\sim$18.5, the
difference between our measurements is the extinction value of 0.14.

\begin{figure}
  \begin{center}
    \includegraphics[width=150mm]{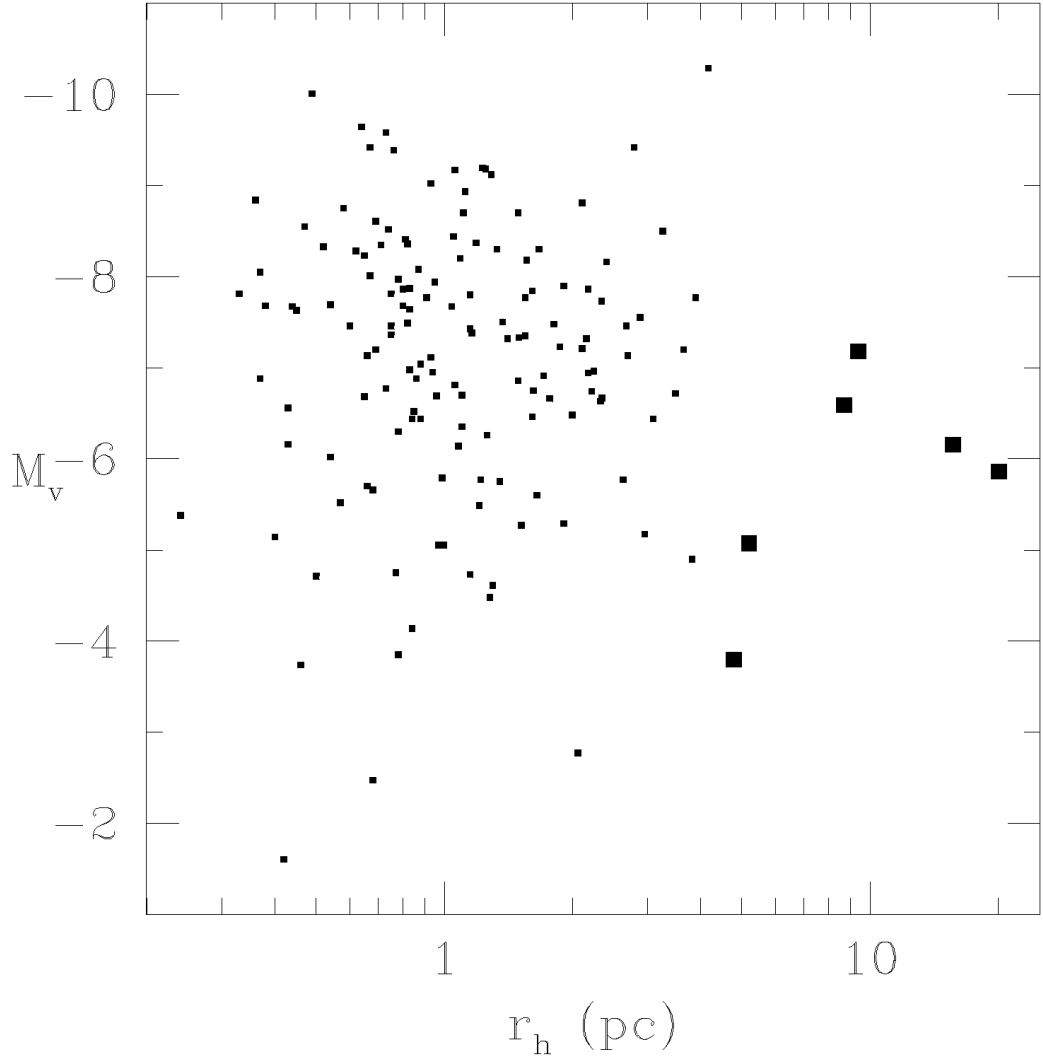}
  \end{center}
  \caption{Absolute magnitudes versus the half-light radius in
    parsecs for both the Milky Way clusters \citep{1996AJ....112.1487H}, shown in
    small boxes, and the six M33 outer halo clusters, shown as larger boxes.}
  \label{mv_rh}
\end{figure}

Our crude estimate for the cluster magnitude limit is
currently $g'_{lim}$ $\approx 20$ ($M_{g} \approx -4.8$).  We
will quantify this more accurately in an upcoming paper by inserting
fake clusters and testing recovery rates using our search methods.  Although our current search limit is comfortably faint, there are
small numbers of still less luminous clusters known to exist in the
Milky Way, for example (the faintest, sparse Palomar-type objects; see
 Figure \ref{mv_rh}).  We can therefore place no quantitative
limits on the numbers of such objects yet to be found in M33.  Note that cluster D \citep{2009ApJ...698L..77H} is a magnitude fainter
than our estimated limit, but was found with $HST$ imaging.  We would
not expect to recover such a cluster independently with the MegaCam data.

\begin{figure}
  \begin{center}
    \includegraphics[width=65mm]{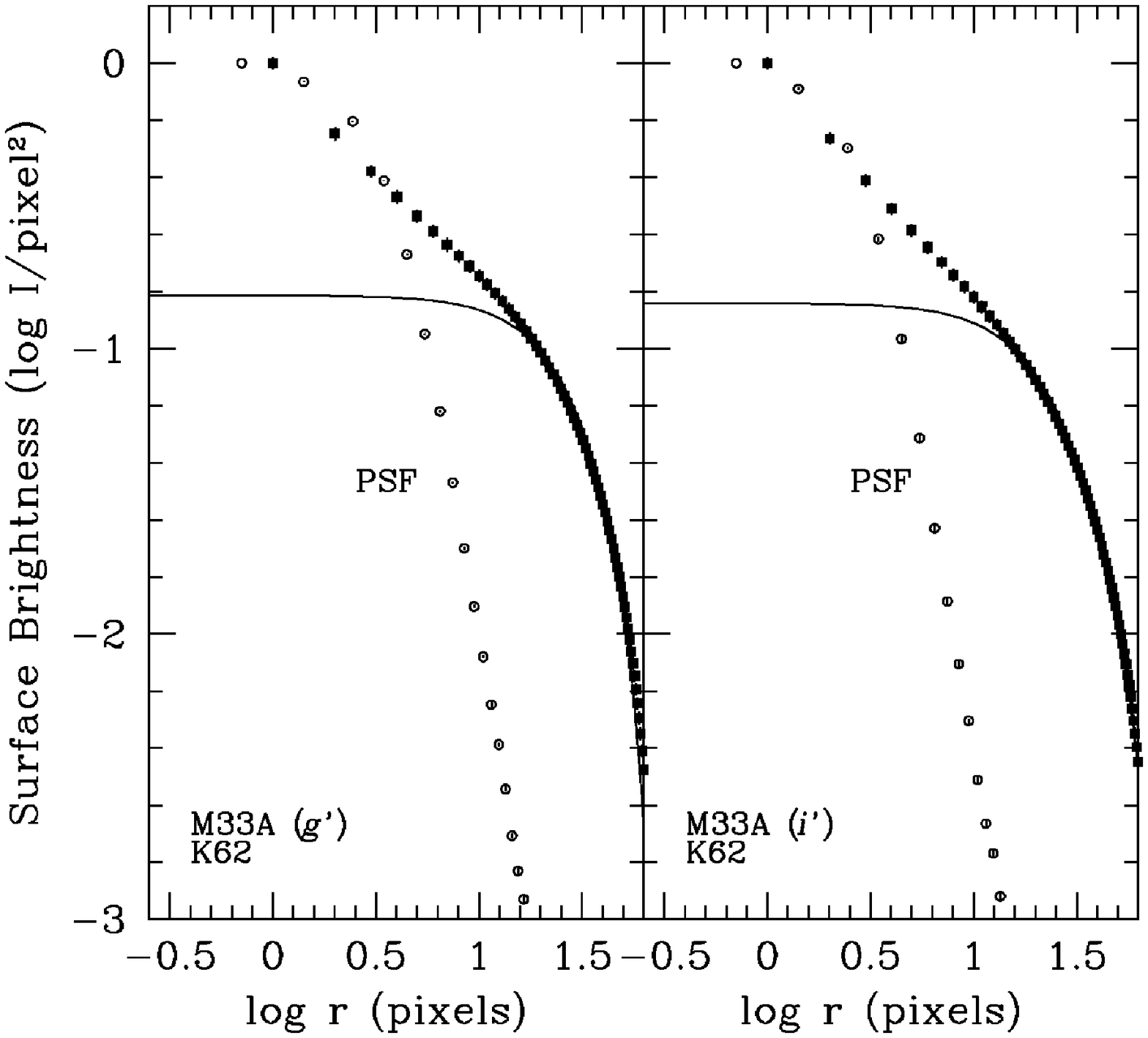}
    \includegraphics[width=65mm]{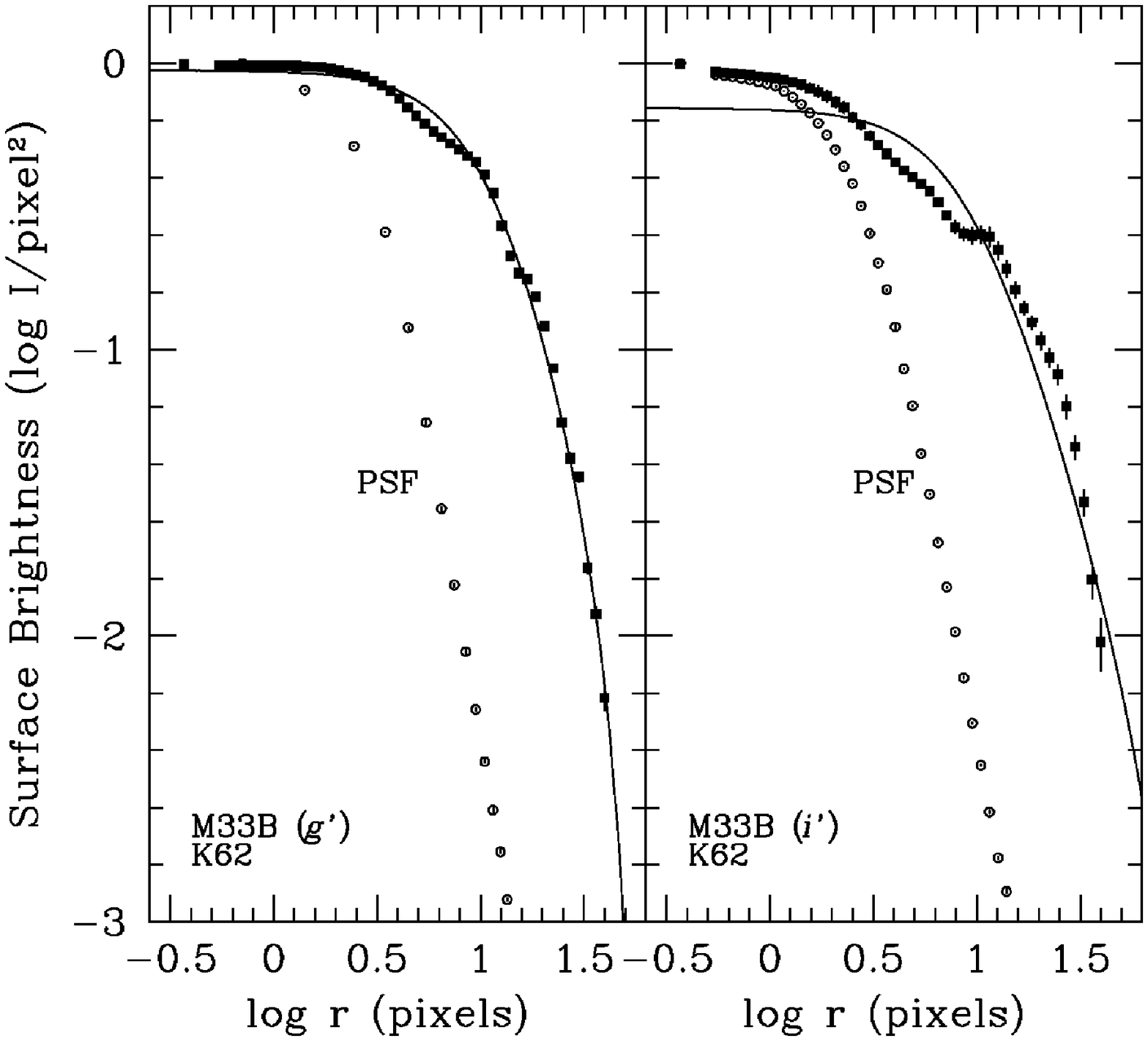}
    \includegraphics[width=65mm]{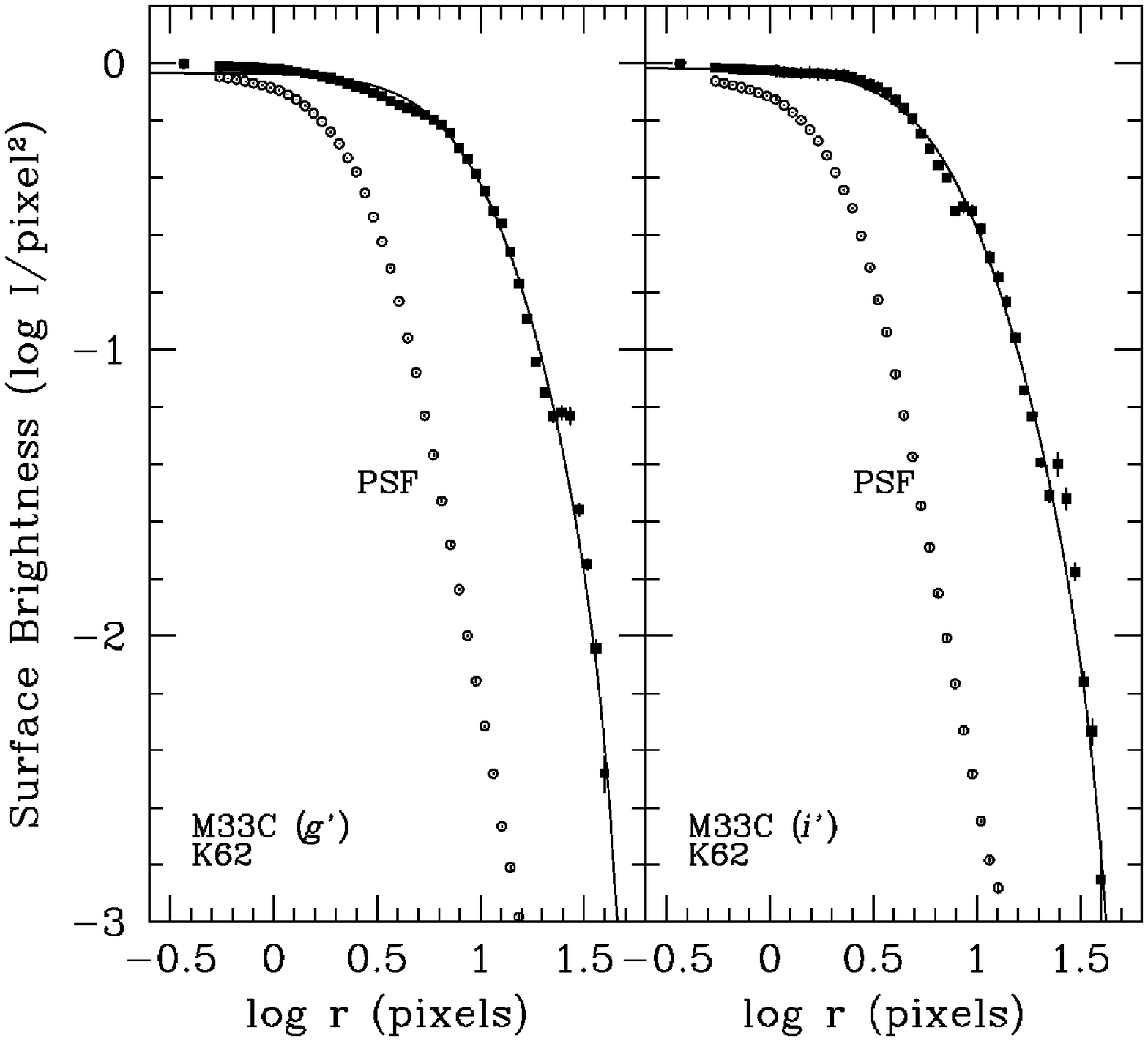}
    \includegraphics[width=65mm]{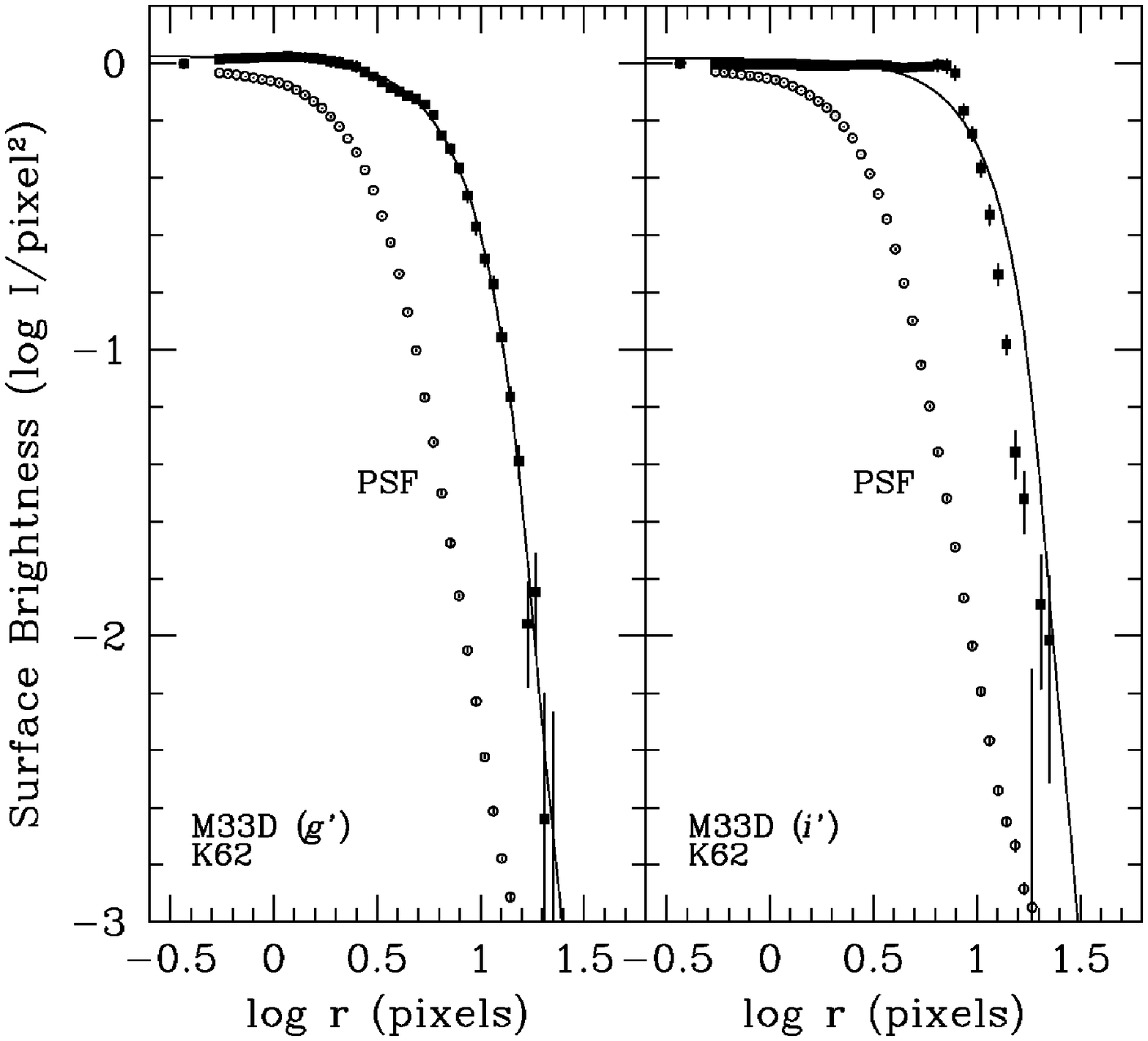}
    \includegraphics[width=65mm]{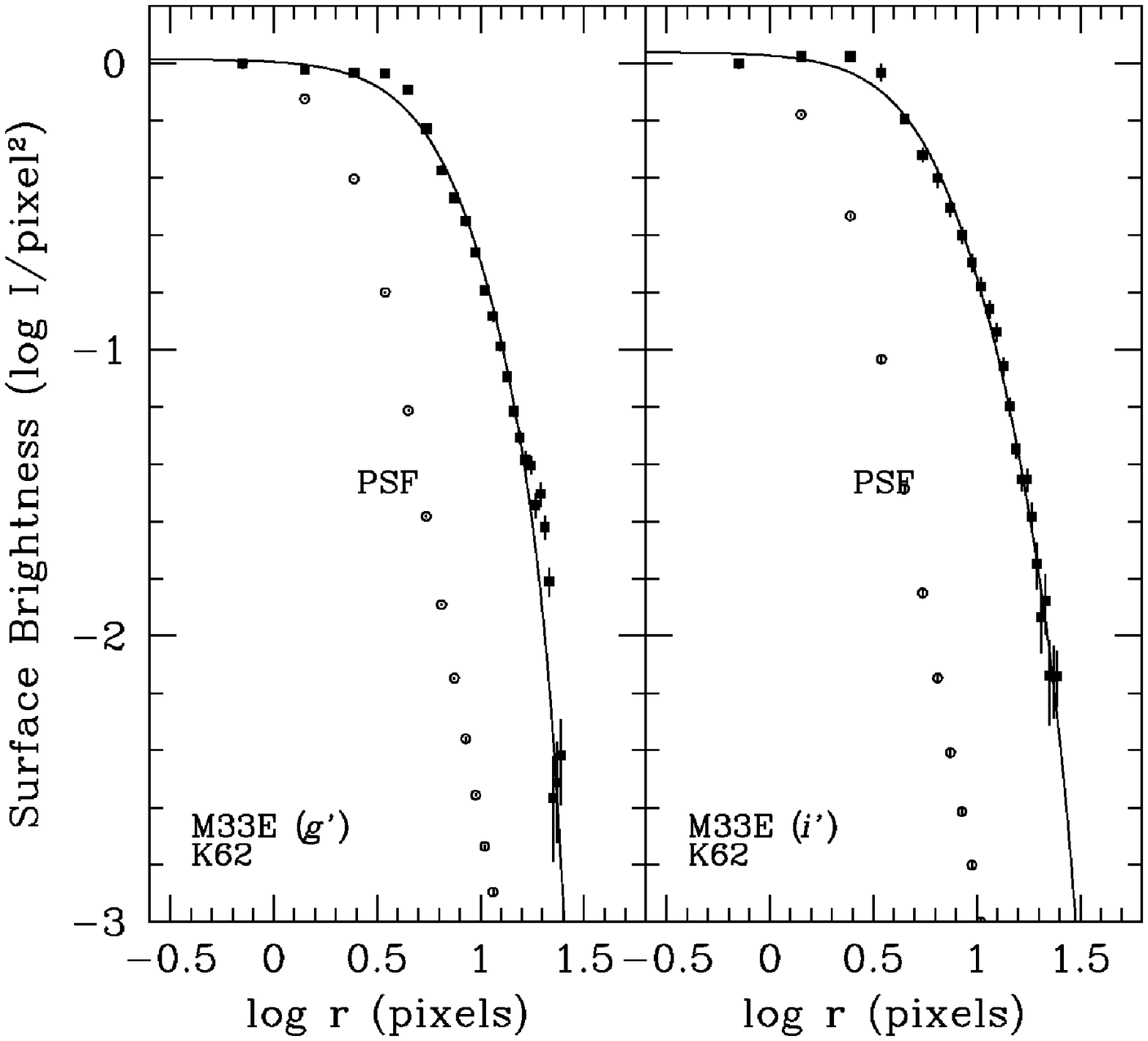}
    \includegraphics[width=65mm]{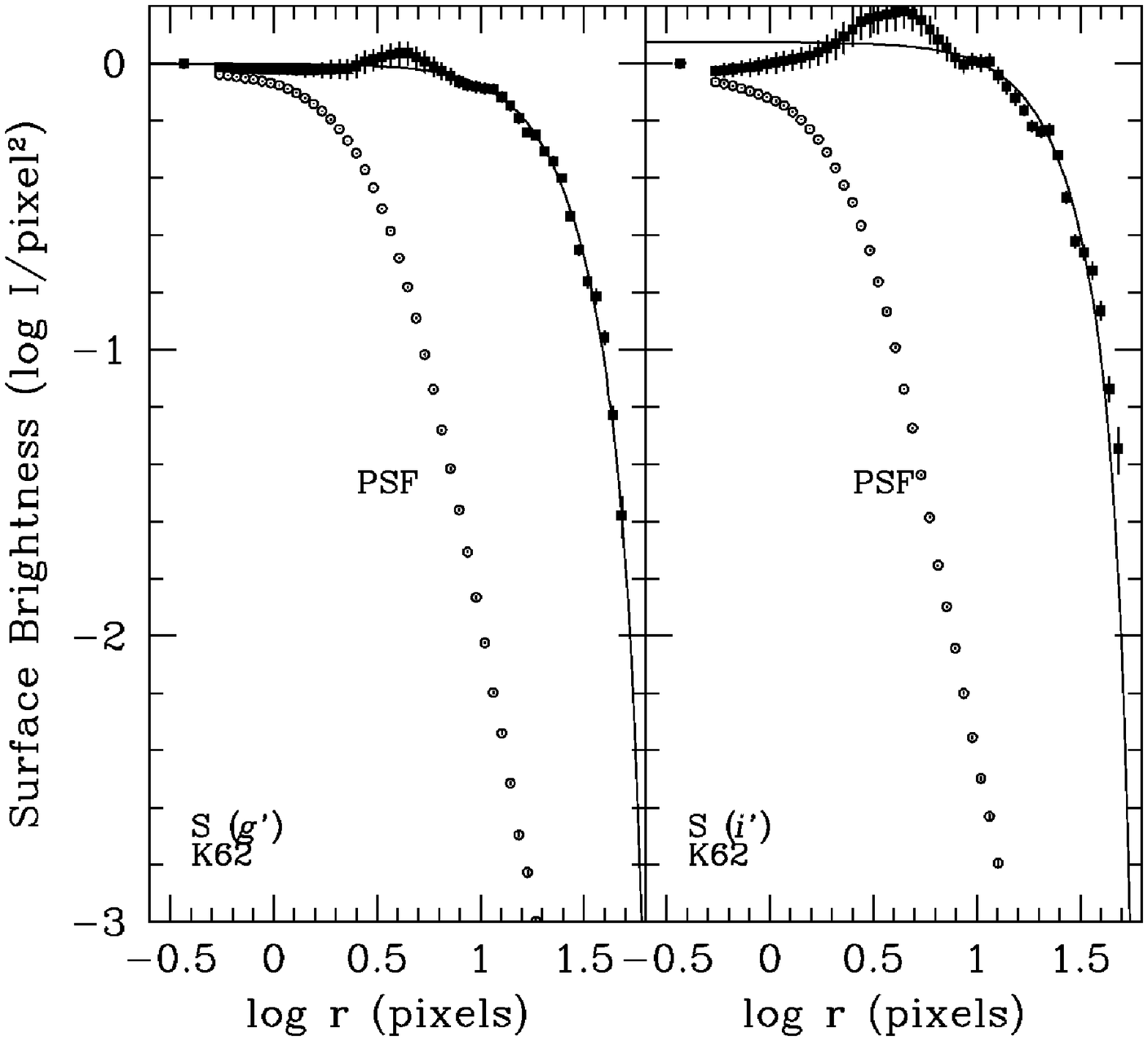}
  \end{center}
  \caption{Examples of radial profiles for each of the six outer halo
  clusters.  The solid points indicate the data, the line is the
  bestfit, the open points show the profile for the PSF.  Note that for clusters A and
  D, the fits did not converge to a simple King-type model solution adequately.}
  \label{radprofs}
\end{figure}

Finally, we measured the structural parameters of all six outer clusters,
including the concentration parameter, and core, half-light and tidal
radii.  We use the GRIDFIT code described by \cite{2008MNRAS.384..563M}, which fits
various King-type cluster models convolved with the measured PSF to each
object.  Here we attempt to fit \cite{1962AJ.....67..471K}, \cite{1966AJ.....71...64K} and \cite{1975AJ.....80..175W}
models to each object.  We also use the KFIT2D code of
\cite{2002AJ....124.2615L} with the \cite{1966AJ.....71...64K} model
as an independent measure.  The results of all fits are shown in Table
\ref{clusterradii}, and examples of the fits are shown in Figures
\ref{radprofs}.  We also include an independent
measurement of the half-light radii using the curve of growth of the
clusters (r$_{ap}$).  For A and D, not all of the three
models converged to successful fits, but the other four clusters gave
high consistency among themselves for their radii.  Fitting models to cluster A was
not successful because of its diffuse nature, while cluster D was
extremely small.  We also note that cluster S has an unusual feature
in its surface profile; \cite{2008AJ....135.1482S} also note that this
cluster is asymmetrical in its inner regions.  Comparing our measured
structural parameters for cluster S with those from
\cite{2008AJ....135.1482S} we similarly find that this cluster has a
very large core radius, although we measure
slightly smaller quantities for each radius - even adjusting the value for
our assumptions for the distance to M33.  

In Figure \ref{mv_rh},
we show the locations of all six M33 halo clusters in luminosity versus
$r_h$, compared with the Milky Way GCs.  All clusters have low
concentrations, similar to the Palomar outer halo clusters in the
Milky Way.  Their half-light radii range from 4 to 20 pc, all larger than the typical mean $r_h \sim 3$ pc for the standard Milky Way clusters, but placing them
in a similar range as many of the outer halo Milky Way clusters.  A full comparison of all M33
clusters will be done in a future paper.

%%%%%%%%%%%%%%%%%%%%%%%%%%%%%%%%%%%%%%%%%%%%%%%%%%%%%%%%%%%%%%%%%%%

\section{Discussion and Conclusions}
\label{conclusion}

\begin{figure*}
  \begin{center}
    \includegraphics[width=80mm]{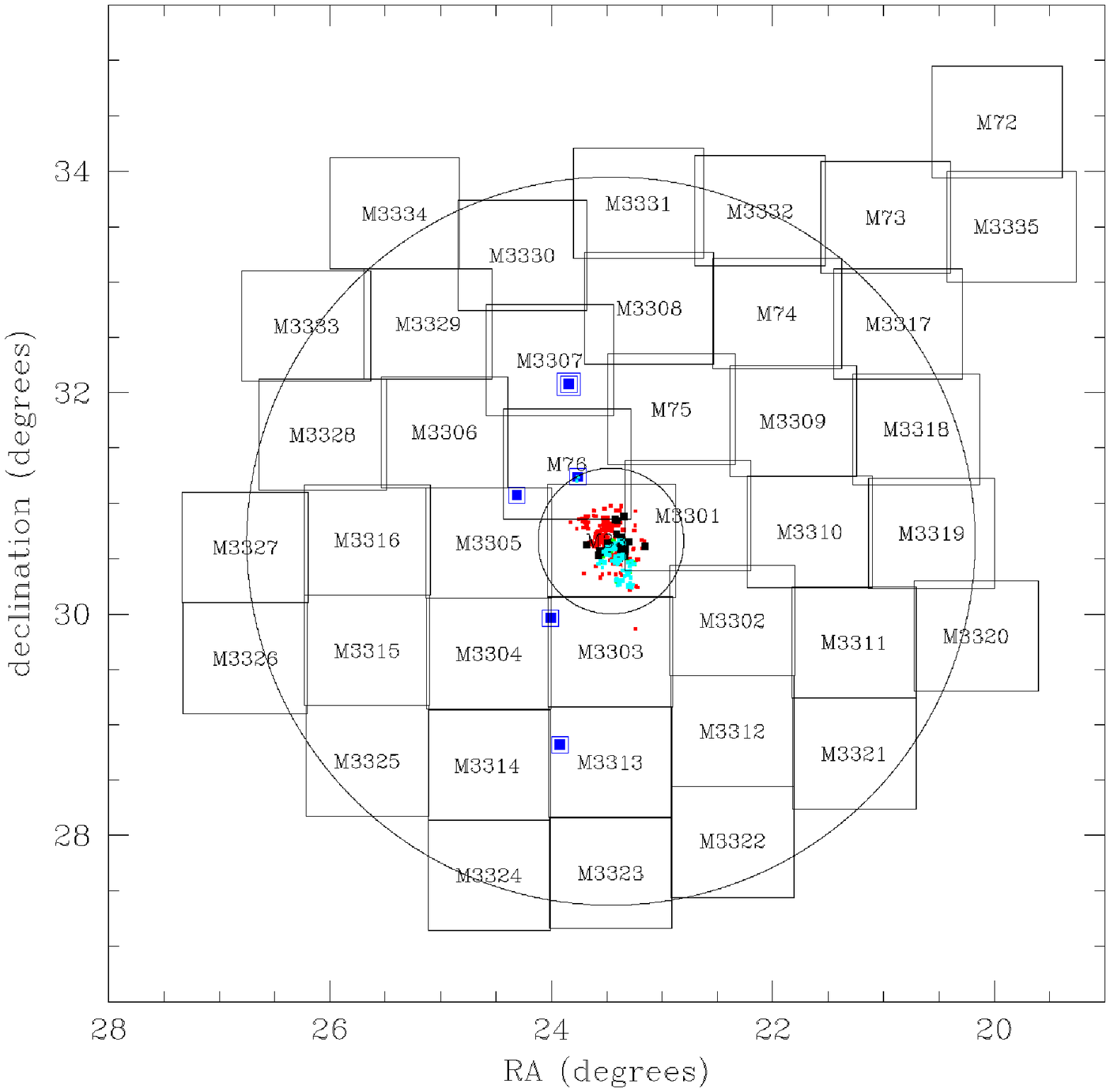}
    \includegraphics[width=80mm]{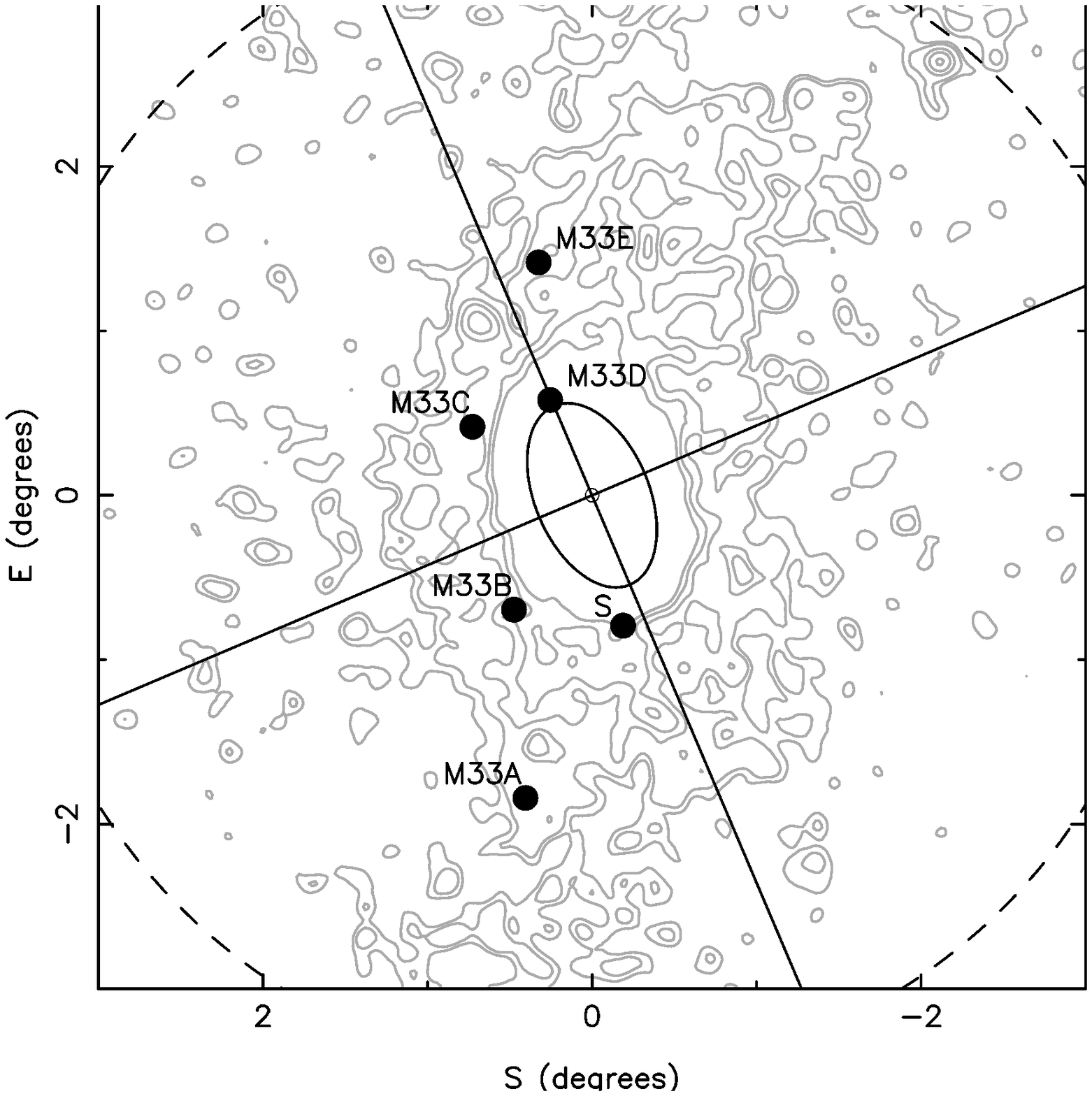}
  \end{center}
  \caption{(Left) The 41 PAndAS frames by CFHT/MegaCam around M33 used in
    this paper.  Each
    square represents the one square-degree field-of-view of MegaCam.
    M33c is the central field.  The two circles represent projected radii of 10
    kpc and 50 kpc centered on M33.  Also shown are the locations of the high-confidence clusters in
    the \cite{2007AJ....134..447S} catalogue (the
    outer halo cluster found by \cite{2008AJ....135.1482S} being the
    only catalogue's object outside the 10 kpc radius), the four outer halo
    clusters in \cite{2009ApJ...698L..77H} enclosed in one box and the
    newly discovered outer halo cluster enclosed in a double box.  The
    current (October, 2010) online \citeauthor{2007AJ....134..447S} catalogue is further subdivided to show
    the location of the 296 original clusters in red from
    \cite{2007AJ....134..447S} in addition to the
    subsequently-discovered 32 clusters in black from
    \cite{2007AJ....134.2168P}, 115 clusters in cyan from
    \cite{2009AcA....59...47Z}, and 161 (115 new) clusters in green from
    \cite{2009ApJ...699..839S}.  There is much overlap between the
    clusters found by \cite{2009AcA....59...47Z} and
    \cite{2009ApJ...699..839S}.  (Right) The outer halo clusters
    overlaid on the substructure map from Figure 13 in
    \cite{2010ApJ...723.1038M}.
}
  \label{locations}
\end{figure*}

We search more than 40 square degrees of the halo of M33 with
CFHT/MegaCam data for outer halo clusters using
both an automated search and visual inspection.  Unexpectedly, we find
only one new cluster, which is smaller, fainter and slightly redder
than the three INT clusters found by \cite{2009ApJ...698L..77H} and
the cluster found by \cite{2008AJ....135.1482S}.  However, it does lie
within the INT area of \cite{2009ApJ...698L..77H} but the object was
not previously recognized due to its small size and faint luminosity.  At a projected
radius of 22 kpc, the new cluster has $g'\approx19.9$,
$(g'-i')\approx0.6$, concentration parameter $c\approx1.0$, a core
radius $r_{c}\approx3.5$ pc, and a half-light radius $r_{h}\approx5.5$
pc.  Its projected location is close to the feature observed in the stellar
substructure (see Figure \ref{locations} and
\citealt{2010ApJ...723.1038M}).  \cite{2009ApJ...698L..77H} note that
the mean color of the previously-known outer halo clusters is
slightly redder ($(V-I)_0$ = 0.88 $\pm$ 0.05 mag) than the inner clusters
($(V-I)_0$ = 0.67 $\pm$ 0.30 mag). Our new cluster is redder still by
$\sim$0.2 mag.

M33 has only six definite outer halo clusters between projected radii of 9 kpc $\leq$ r $\leq$ 50 kpc and to $g'_{lim}$ $\approx 20$. We also find 2440 cluster candidates of various degrees of confidence, and although the vast majority are likely to be
background contaminants, at least some of the $\sim 90$
highest-confidence candidate objects beyond the M33 disk may be faint
but genuine clusters. We cannot yet assume all the highest-confidence
candidate objects are clusters without further confirmation.  We will
use IR data (now being acquired) and structural parameters in an upcoming paper to
determine this more securely.  

How many clusters could we expect to find in M33?  M31 has 67 outer
halo clusters already discovered, 61 of which lie in the PAndAS
footprint that has been analyzed so far.  These clusters have
comparable luminosity to the M33 outer halo clusters and are located
at projected radius 30 kpc $\leq r \leq$ 130 kpc
\citep{2010ApJ...717L..11M, 2011Huxorinprep}. \cite{2009ApJ...698L..77H} found a GC surface density of $\sim$0.4
deg$^{-2}$ with their 12 deg$^2$ study, which they note is about half
that derived for M31 over the radial range 30kpc $\leq r \leq$ 130
kpc.  Here we find an even lower GC surface density of 0.15
deg$^{-2}$.  We note that the search in M31's outer halo is not yet complete so its GC
surface density is likely to increase.  M33 appears to therefore lack this type of
cluster.  We briefly mention two scenarios that could have resulted in this observed
difference.  

M33 could have had a different accretion history
compared to M31 - a conclusion that has
been drawn before from studies of the inner regions
\citep{2010arXiv1007.1042S}, but is now also indicated by the outer
halo data.  If M33 never interacted with M31 before, M33 would have
had a dramatically less active accretion history.  

The most compelling evidence for an accretion origin for the
outer halo clusters comes from the Sagittarius dwarf in the MW \citep{1995MNRAS.277..781I}, and from the GCs and tidal debris streams in M31 \citep{2010ApJ...717L..11M}, but it is still far from clear how general a result this is."

However, another exciting and
more likely prospect, given the tidal distortion of M33, is that perhaps some of M33's outer halo
clusters were heavily stripped off in a previous dynamical interaction
with M31 \citep{2009ApJ...698L..77H, 2010arXiv1007.1042S}.  Some of the GCs originally belonging to M33 may
now be closer to M31, but it will be difficult to disentangle the
populations.  A more detailed
comparison will require spectroscopic studies of these clusters to
determine properties that may link the divided populations.  Although
unlikely, some clusters may be beyond the area that we have imaged so
far around M33.  These scenarios are not mutually exclusive.  Further discussion and
comparison with the M33 halo star population will come in subsequent
work now in progress.

%%%%%%%%%%%%%%%%%%%%%%%%%%%%%%%%%%%%%%%%%%%%%%%%%%%%%%%%%%%%%%%%%%%

\acknowledgements
RC and WEH thank the Natural Sciences and Engineering Research Council
of Canada for financial support.  This research has made use of the
NASA/IPAC Extragalactic Database (NED) which is operated by the Jet
Propulsion Laboratory, California Institute of Technology, under
contract with the National Aeronautics and Space Administration.

Based on observations obtained with MegaPrime/MegaCam, a joint project
of CFHT and CEA/DAPNIA, at the Canada-France-Hawaii Telescope (CFHT)
which is operated by the National Research Council (NRC) of Canada,
the Institute National des Sciences de l'Univers of the Centre
National de la Recherche Scientifique of France, and the University of
Hawaii.  RC would like to thank the CFHT staff for much support.

Thanks to all our collaborators in the PAndAS and to the anonymous
referee for providing useful
comments that improved this paper.  RC would also like to
thank the Universities of Edinburgh and Cambridge for funding collaborative visits.

%%%%%%%%%%%%%%%%%%%%%%%%%%%%%%%%%%%%%%%%%%%%%%%%%%%%%%%%%%%%%%%%%%%

\bibliography{adsbibliography.bib}

%%%%% Tables %%%%%%%%%%%%%%%%%%%%%%%%%%%%%%%%%%%%%%%%%%%%%%%%%%%%%%

\begin{deluxetable}{ccc}
  \tablecaption{Source Extractor values used on the MegaCam images. \label{separams}}
  \tabletypesize{\scriptsize}
  \tablewidth{0pt}
  \tablehead
  {
    \colhead{Parameter}&\colhead{Default Value}&\colhead{New Value}
  }
  \startdata
 DETECT\_MINAREA&5&4\\
 THRESH\_TYPE&-&RELATIVE\\
 DETECT\_THRESH&1.5&5\\
 DEBLEND\_NTHRESH&32&8\\
 CLEAN\_PARAM&1.0&1.5\\
 PHOT\_APERTURES&5&3\\
 PHOT\_AUTOPARAMS&2.5,3.5&2.0,2.5\\
 PHOT\_PETROPARAMS&2.0,3.5&2.0,2.5\\
 PHOT\_FLUXFRAC&-&0.5\\
 SATUR\_LEVEL&50000&60000\\
 MAG\_ZEROPOINT&0.0&$g'$=26.7,$i'$=25.98\\
 BACKPHOTO\_TYPE&GLOBAL&LOCAL\\
  \enddata
\end{deluxetable}

\begin{deluxetable}{cccccccccc}
  \tablecaption{Categorized objects in M33 PAndAS frames.  The column
    headers are as follows: SM1 are the \cite{2007AJ....134..447S}
    catalogue's confirmed clusters,  OH are the outer halo clusters, 1
    are our highest-confidence clusters, 2 are the possible clusters,
    3 are the background galaxies, 4 are unknown objects and 5 are
    stellar objects.  The numbers in brackets after the highest-confidence
    and possible clusters indicate those candidates which matched
    objects in SM1.   \label{objects}}
  \tabletypesize{\scriptsize}
  \tablewidth{0pt}
  \tablehead
  {
    \colhead{Frame}&\colhead{SM1}&\colhead{OH}&\colhead{1}&\colhead{2}&\colhead{3}&\colhead{4}&\colhead{5}
  }
  \startdata
  M3301&34&-&7(3)&354(16)&106&114&0\\
  M3302&-&-&1&69&56&8&27\\
  M3303&1&1$^{c,e}+1$&3&96&43&16&8\\
  M3304&-&1$^{d,e}$&2&69&86&16&7\\
  M3305&-&1$^{b,e}$&2&83&51&19&5\\
  M3306&-&-&2&49&92&20&0\\
  M3307&-&1$^{g}$&2&31&79&16&0\\
  M3308&-&-&0&45&149&13&0\\
  M3309&-&-&0&77&104&11&0\\
  M3310&-&-&0&64&77&7&0\\
  M3311&-&-&1&27&74&6&0\\
  M3312&-&-&3&38&96&11&0\\
  M3313&-&1$^e$&2&64&60&13&0\\
  M3314&-&-&3&44&55&13&1\\
  M3315&-&-&2&35&87&31&0\\
  M3316&-&-&0&56&81&14&0\\
  M3317&-&-&1&44&114&9&0\\
  M3318&-&-&0&25&121&5&0\\
  M3319&-&-&0&55&96&16&0\\
  M3320&-&-&0&29&86&6&0\\
  M3321&-&-&0&8&74&9&0\\
  M3322&-&-&3&26&66&11&1\\
  M3323&-&-&0&18&69&6&1\\
  M3324&-&-&1&63&61&16&3\\
  M3325&-&-&0&31&61&15&3\\
  M3326&-&-&1&29&104&15&0\\
  M3327&-&-&0&24&81&14&3\\
  M3328&-&-&1&32&115&14&3\\
  M3329&-&-&5&72&81&19&6\\
  M3330&-&-&5&67&117&15&0\\
  M3331&-&-&2&61&127&25&0\\
  M3332&-&-&1&58&148&22&6\\
  M3333&-&-&3&39&129&16&2\\
  M3334&-&-&2&35&116&15&2\\
  M3335&-&-&2&29&87&25&0\\
  M72&-&-&2&28&71&11&1\\
  M73&-&-&1&45&93&13&2\\
  M74&-&-&0&52&61&12&1\\
  M75&-&-&0&43&95&22&0\\
  M76&19&2$^{a,e}$&32(8)&234(2)&54&86&3\\
  M33c&374$^f$&-&259(95)&1521(138)&84&954&19\\
\multicolumn{10}{l}{$^a$ Also found in M3305}\\
\multicolumn{10}{l}{$^b$ Also found in M76}\\
\multicolumn{10}{l}{$^c$ Also found in M3304}\\
\multicolumn{10}{l}{$^d$ Also found in M3303}\\
\multicolumn{10}{l}{$^e$ Also found by \cite{2009ApJ...698L..77H}}\\
\multicolumn{10}{l}{$^f$ Does not include those on frames M3301, M3303
  or M76)}\\
\multicolumn{10}{l}{$^g$ New cluster identified in this paper.}\\
  \enddata
\end{deluxetable}

\begin{deluxetable}{ccccccccc}
  \tablecaption{Outer halo cluster positions, luminosities and
    colors.  We assume a distance of 870 kpc to M33, consistent with SM07 and
    \cite{2009ApJ...698L..77H}, and that M33's center is located at
    (01$^h$33$^m$50.9$^s$, 30$^d$39$^m$37$^s$). \label{clusterlums}}
  \tabletypesize{\scriptsize}
  \tablewidth{0pt}
  \tablehead
  {
    \colhead{}&\multicolumn{2}{c}{Degrees}&\multicolumn{2}{c}{Galactocentric distance}&\colhead{ }&\colhead{ }&\colhead{ }&\colhead{ }\\ 
    \colhead{Cluster}&\colhead{RA}&\colhead{Dec}&\colhead{arcmins}&\colhead{kpc}&\colhead{$g'_0$}&\colhead{$(g'-i')_0$}&\colhead{$V_0$}&\colhead{$(V-I)_0$}\\
  }
  \startdata
  A&23.92388&28.82086&112&28.4&19.1&0.7&18.8&0.9\\
  B&24.00865&29.96372&48&12.2&17.8&0.7&17.5&0.8\\
  C&24.31026&31.07433&45&11.3&18.4&0.8&18.1&0.8\\
  D&23.75916&31.23925&37&9.4&21.3&0.8&20.9&1.0\\
  E&23.84466&32.07559&87&21.9&19.8&0.5&19.6&1.1\\
  S&23.24374&29.8675&49&12.3&18.9&0.8&18.5&0.8\\
  \enddata
\end{deluxetable}

\begin{deluxetable}{ccccccccc}
  \tablecaption{Outer halo cluster structural
    parameters, including concentration, c, core radii, half-light
    radii, and tidal radii, using both GRIDFIT and KFIT2D.  We report
    the best-fit to the data, whether in $g'$ or in $i'$.  Also shown are
    half-light radii estimates (r$_{ap}$) using curves-of-growth, for
    an independent check on the upper limit of the half-light radii.
    We assume a distance of 870 kpc to M33, consistent with SM07 and
    \cite{2009ApJ...698L..77H}, and corresponding to a distance
    modulus of (m-M)$_0$ = 24.69. \label{clusterradii}}
  \tabletypesize{\scriptsize}
  \tablewidth{0pt}
  \tablehead
  {
    \colhead{ }&\colhead{ }&\colhead{ }&\colhead{ }&\colhead{ }&\multicolumn{4}{c}{Radii (pc)}\\ 
    \colhead{Cluster}&\colhead{Model}&\colhead{Band}&\colhead{Seeing/FWHM}&\colhead{c}&\colhead{Core}&\colhead{Half}&\colhead{Tidal}&\colhead{r(ap)}\\
  }
  \startdata
  &K62&i&3.8&0.5&16.9&20.1&68.0&11.7\\
  A&K66&i&3.8&0.4&17.7&20.3&83.2& \\
  &W  &-&-&-&-&-&-& \\
  &kfit2d&i&3.8&0.7&11.7&11.6&59.8& \\
  \hline
  &K62&g&4.0&0.8&6.1&9.4&44.0&9.0\\
  B&K66&g&4.0&0.9&6.3&9.4&56&\\
  &W  &g&4.0&1.0&6.6&9.4&87.3&\\
  &kfit2d&g&4.0&1.1&6.0&10.9&78.5&\\
  \hline
  &K62&g&4.0&0.8&5.6&8.7&40.7&7.8\\
  C&K66&i&3.3&1.0&4.5&7.3&48.1&7.8\\
  &W  &i&3.3&1.0&4.9&7.1&70.8&\\
  &kfit2d&i&3.3&1.3&3.8&8.8&81.9&\\
  \hline
  &K62&g&4.5&0.3&4.7&4.8&13.6&4.7\\
  D&K66&-&-&-&-&-&-&\\
  &W  &-&-&-&-&-&-&\\
  &kfit2d&g&4.5&0.4&4.1&3.7&9.8&\\
  \hline
  &K62&i&2.6&0.8&3.3&5.2&25.5&5.1\\
  E&K66&i&2.6&0.9&3.5&5.2&31.4&\\
  &W  &g&3.1&1.2&3.6&5.6&65.8&5.1\\
  &kfit2d&i&2.6&1.1&2.7&5.0&32.4&\\
  \hline
  &K62&g&4.6&0.3&15.8&15.7&42.7&18.7\\
  S&K66&i&3.4&0.4&14.8&16.7&67.31&18.7\\
  &W&i&3.4&0.2&15.8&19.2&121.4&\\
  &kfit2d&i&3.4&0.7&17.9&18.7&93.7&\\
  \enddata
\end{deluxetable} 

%%%%% Figures %%%%%%%%%%%%%%%%%%%%%%%%%%%%%%%%%%%%%%%%%%%%%%%%%%%%%

%%%%%%%%%%%%%%%%%%%%%%%%%%%%%%%%%%%%%%%%%%%%%%%%%%%%%%%%%%%%%%%%%%%

\end{document}